\documentclass[12pt,preprint]{aastex}
\usepackage{color}
\usepackage{ulem}
\usepackage{subfig}

\newcommand{\al}{\mbox{${}^{26}{\rm Al}$}}
\newcommand{\ca}{\mbox{${}^{41}{\rm Ca}$}}
\newcommand{\fe}{\mbox{${}^{60}{\rm Fe}$}}
\newcommand{\alratio}{\mbox{${}^{26}{\rm Al} / {}^{27}{\rm Al}$}} 
\newcommand{\oxvi}{\mbox{$^{16}{\rm O}$}}
\newcommand{\oxvii}{\mbox{$^{17}{\rm O}$}}
\newcommand{\oxviii}{\mbox{$^{18}{\rm O}$}}
\newcommand{\fxviii}{\mbox{$^{18}{\rm F}$}}
\newcommand{\dxvii} {\mbox{$\delta^{17}{\rm O}$}}
\newcommand{\dxviii}{\mbox{$\delta^{18}{\rm O}$}}
\newcommand{\Msol}{\mbox{$\, \rm{M}_{\odot}$}} 
\newcommand{\promille}{%
  {\small
  \relax\ifmmode\promillezeichen
        \else\leavevmode\(\mathsurround=0pt\promillezeichen\)\fi}  }
\newcommand{\promillezeichen}{%
  \kern-.05em%
  \raise.7ex\hbox{\the\scriptfont0 0}%
  \kern-.15em/\kern-.15em%
  \lower.25ex\hbox{\the\scriptfont0 00}}

\begin{document}

\title{
Collateral Effects on Solar Nebula Oxygen Isotopes due to Injection of ${}^{26}{\rm Al}$ by a Nearby Supernova
}

\author{Carola I. Ellinger\altaffilmark{1}, Patrick A. Young\altaffilmark{2}, \& Steven J.\ Desch\altaffilmark{2}}

\altaffiltext{1}{Department of Physics, Arizona State University, P.O.Box 1504, Tempe, AZ 85287-1504}
\altaffiltext{2}{School of Earth and Space Exploration, Arizona State
U261niversity, Tempe, AZ 85287}


\begin{abstract}

Injection of material from a core-collapse supernova into the solar system's already-formed 
disk is one proposed mechanism for producing the short-lived radionuclides, such as \al\, 
and \ca, 
inferred from isotopic studies of meteorites to have existed in the solar nebula.
This hypothesis has recently been challenged on the basis that the injection of enough supernova
material to match the meteoritic abundances of $\al$ and \ca~ would produce large, measureable, and 
unobserved collateral effects on oxygen isotopes. 
Here we calculate again the shifts in oxygen isotopes due to injection of supernova
material in the solar nebula, using a variety of nucleosynthetic conditions of our own progenitor explosions.  
Unlike previous studies of this type, we also consider the effect of non-homogeneity in abundance distribution 
of the nucleosynthesis products after the explosion. 
We calculate the shifts in oxygen isotopes due to injection of sufficient supernova material to 
produce the meteoritic abundances of $\al$ and \ca, and analyze the predicted shifts in detail 
for compatibility with meteoritic data.
We find that the range in possible isotopic shifts is considerable and sensitive to parameters 
such as progenitor mass and anisotropy of the explosion; however, a small number of compatible 
scenarios do exist.
Because of the wide range of outcomes and the sensitivity of isotopic yields to assumed 
conditions, it is difficult to constrain the supernova that may have led to injection of 
$\al$ in the solar nebula. Conversely, we argue that the existence of viable counterexamples demonstrates that
it is premature to use oxygen isotopes to rule out injection of $\al$ and \ca~ into the solar nebula
protoplanetary disk by a nearby supernova.
\end{abstract} 

%=========================================================================================%
% SECTION 2: INTRODUCTION                                                                 %
%=========================================================================================%

\section{Introduction}

From isotopic studies of meteorites it is known that the solar nebula
contained at least a dozen different short-lived radionuclides, or SLRs
\citep[see reviews by][]{MckeD03, MeyeZ06, Wadh07}.
Identification of the sources of these SLRs could greatly constrain the 
Sun's birth environment and processes acting during star formation. 
The half-lives of some of these isotopes are shorter than the timescales
$\sim 10^{6} - 10^{7} \, {\rm yr}$ typically associated with star formation, 
so they must have been produced near the time and place of the Sun's formation.
The SLRs $\al$, $\ca$ and $\fe$, in particular, cannot have been inherited from
the Sun's molecular cloud in abundances consistent with ongoing Galactic nucleosynthesis,
and must have been late additions \citep{Jaco05}. 
A leading candidate for the source of these and other SLRs is one or more core-collapse 
supernovae in the Sun's birth environment, contaminating either its molecular cloud 
\citep{CameT77, VanhB02, GounK09}, or
its protoplanetary disk \citep{Chev00, OuelD05, Ouel07, OuelDH10, LoonT06}.
Another leading candidate is production of SLRs by irradiation (by solar cosmic 
rays, essentially), within the solar nebula \citep{Goun01}.
Because this latter mechanism by itself is inadequate to explain the abundance of 
$\fe$ in the early solar system \citep{LeyaH03, Goun06}, it is generally
accepted that the source of $\fe$ is core-collapse supernovae \citep[e.g.][]{Wadh07}, 
although it is not clear whether the source of $\fe$ is a single, nearby 
supernova, or many (possibly distant) supernovae \citep[as in][]{GounK09}.

The origins of the other SLRs are also debated.
A correlation between $\al$ and $\ca$ has been observed in meteorites, demanding
a common source for these two isotopes {\it after} the formation of the solar nebula,
as in the so called ``late-injection" hypothesis of \citet{SahiG98}.
It is not yet clear whether these two SLRs are correlated with $\fe$.
If evidence for a corelation could be found, this would strongly suggest that $\al$ 
and $\ca$ were injected by the same supernova or supernovae that injected $\fe$.
The lack of such evidence, though, leaves open the possibility that $\al$ and $\ca$ 
were created by irradiation within the solar nebula while $\fe$ was injected separately
by one or more supernovae, into the Sun's molecular cloud or protoplanetary disk. 

The abundances of the SLRs alone have not yet enabled a discrimination between 
these possibilities, but \citet[][hereafter GM07]{Goun07} have proposed 
that the oxygen isotopic ratios of early solar system materials may be used to rule 
out certain hypotheses.  
Specifically, they argue that if $\al$ and $\ca$ were injected by a nearby
supernova into the Sun's protoplanetary disk, sufficient to produce the observed 
meteoritic ratio $\alratio \approx 5 \times 10^{-5}$ \citep{MacPD95}, then 
the oxygen isotopic ratio of the solar nebula would be considerably altered:
solar nebula materials formed before the injection would have oxygen isotopic 
ratios {\it significantly} different from later-formed materials. 
GM07 calculated the shifts in oxygen isotopic ratios accompanying injection of 
supernova $\al$ into the Sun's protoplanetary disk, using the isotopic yields in
bulk supernova ejecta calculated by \citet{Raus02}.
A robust prediction of the GM07 models is that ${}^{17}{\rm O} / {}^{16}{\rm O}$ 
of pre-injection materials should be significantly higher, by several percent,
than post-injection materials.
Examples of pre-injection materials may exist in meteorites, or especially in the 
solar wind sample returned by the ${\it Genesis}$ mission \citep{Burn03}. 
Since preliminary results from {\it Genesis} suggest the Sun is {\it not} 
isotopically heavy in oxygen \citep{Mcke09}, and because no such  
${}^{17}{\rm O}$-rich (or ${}^{16}{\rm O}$-poor) components have been discovered 
in meteorites, GM07 rule out a supernova origin for the $\al$ and $\ca$ in meteorites.

The purpose of this paper is to reproduce and refine the method pioneered by GM07, 
and to test the conclusion that $\al$ cannot have a supernova origin.
GM07 originally considered only bulk ejecta of spherically symmetric supernova 
explosions.
We begin our analysis with this case, but make necessary refinements to the method,
and use our current nucleosynthesis models to predict the isotopic yields.
We then expand on the analysis of GM07, calculating the isotopic yields  by allowing 
the disk to intercept ejecta from different 
parts of the supernova explosion rather than a uniformly mixed total yield, and by examining 
anisotropic explosions.
We also simultaneously consider the injection of ${}^{41}{\rm Ca}$ into the disk.

The paper is organized as follows. 
In \S 2, we outline the method used to calculate shifts in oxygen isotopic
composition due to supernova injection of $\al$ and $\ca$, including updates to the method of GM07.
In \S 3 we describe the results of nucleosynthesis simulations we have carried out,
to determine the isotopic yields in supernova ejecta under various explosion scenarios.
We determine the inputs needed to compute the shifts in solar nebula oxygen isotopic 
composition.  
These shifts in oxygen isotope before and after injection are presented in \S 4, and 
in \S 5 we draw conclusions.

%=========================================================================================%
% SECTION 2: METHOD                                                                       %
%=========================================================================================%

\section{Method}

\subsection{Calculation of Isotopic Shifts} 

The method of GM07 is fairly straightforward. 
They assume that meteoritic components that sample the solar nebula's starting composition,
{\it before} the acquisition of ${}^{26}{\rm Al}$, can be identified and measured. 
Likewise, they assume samples {\it after} the acquisition of ${}^{26}{\rm Al}$ can be 
identified and measured. 
Any difference in the oxygen isotopic content between samples of those two 
groups would then constitute a shift in oxygen isotopes brought about by the injection of 
supernova material.
The preditced shift in oxygen isotopes due to injection of supernova material into the 
protoplanetary disk can then be compared to the actual difference in oxygen isotopes
before and after.
In practice, because the vast majority of meteoritic components sample the solar nebula
after injection, GM07 assumed a ``final" value for the solar nebula oxygen isotopes, and
used the isotopic yields in supernova ejecta to predict the initial composition.
Testing the supernova injection hypothesis thus amounts to finding meteoritic inclusions 
with this initial oxygen isotopic composition. 
Such inclusions should have evidence for no live ${}^{26}{\rm Al}$ at the time of their
formation, and should be among the oldest meteoritic inclusions. 

The earliest-formed solids in the solar system are widely accepted to be the calcium-rich, 
aluminum-rich inclusions (CAIs), both because they contain minerals that are the first 
solids expected to condense in a cooling solar nebula \citep{Gros72}, and because their
Pb-Pb ages are the oldest measured, at 4568.6 Myr \citep{BouvW09}.
It is worth noting that because many of the minerals in CAIs are condensates, their
isotopic composition should reflect that of the solar nebula gas. 
The vast majority of CAIs have inferred initial ratios $\alratio \approx 5 \times 10^{-5}$
or appear to have been isotopically reset at a later date \citep{MacPD95}.
Only in a handful of CAIs known as ``FUN" CAIs (fractionation with unknown nuclear
effects) has it been possible to set firm upper limits on the initial $\alratio$ ratio
and show these CAIs did not contain live $\al$ when they formed \citep{FaheG87, MacPD95}.
Thus, CAIs overall reflect the composition of the solar nebula at an early time, and 
FUN CAIs possibly record the oxygen isotopic abundance before the solar nebula acquired $\al$.

To make more precise statements, it is necessary to quantify the oxygen isotopic composition
of the nebula and various components. 
The molar fraction of oxygen in gas and rock can vary, so the relevant quantities are the 
ratios of the stable oxygen isotopes, $\oxvii / \oxvi$ and $\oxviii / \oxvi$.
In the field of cosmochemistry, these ratios are commonly expressed as deviations from a 
standard, in this case Standard Mean Ocean Water (SMOW), which has 
$\oxvii / \oxvi = 3.8288 \times 10^{-4}$ and $\oxviii / \oxvi = 2.0052 \times 10^{-3}$
\citep{oneil86}. 
The fractional deviations of the isotopic ratios from these standard values are
$\dxvii$ and $\dxviii$, and are measured in parts per thousand, or ``permil" ($\promille$). 
[That is, $\dxvii = 1000 \times \left( (\oxvii / \oxvi) / (\oxvii / \oxvi)_{\rm SMOW} - 1 \right)$.]
It is also standard to report the quantity $\Delta^{17}{\rm O} \approx \dxvii - 0.52 \, \dxviii$,
because this quantity is conserved during almost all chemical fractionation processes. 
[More precisely, $\Delta\oxvii \equiv \ln(1+\dxvii) - 0.5247 \ln(1+\dxviii)$ \citep{Mill02}.]

It is clear that the final oxygen isotopic composition of the nebula, $(\dxvii,\dxviii)'$, 
will depend on its starting composition $(\dxvii,\dxviii)_{0}$, the composition of the 
supernova material, $(\dxvii,\dxviii)_{\rm SN}$, and the mass of supernova material injected
(relative to the mass of the disk). 
It is straightforward to show that 
\begin{equation}
\delta^{17}{\rm O}' - \delta^{17}{\rm O}_{0} =
 \frac{ x }{ 1 + x } \, \left( \delta^{17}{\rm O}_{\rm SN} - \delta^{17}{\rm O}_{0} \right),
 \label{eq:shift}
\end{equation}
where $\delta^{17}{\rm O}_{\rm SN}$ is the isotopic ratio of the supernova material
injected into the disk, and $x \equiv M({}^{16}{\rm O})_{\rm SN} /  M({}^{16}{\rm O})_{\rm disk}$
measures the mass of injected oxygen relative to the oxygen present in the disk (with a similar
formula applying to $\dxviii'$).
In terms of the masses involved, 
\begin{equation}
x = \frac{ M({}^{16}{\rm O})_{\rm SN}   }{ M({}^{26}{\rm Al})_{\rm SN}   } \times
    \frac{ M({}^{26}{\rm Al})_{\rm SN}  }{ M({}^{26}{\rm Al})_{\rm disk} } \times
    \frac{ M({}^{26}{\rm Al})_{\rm disk}}{ M({}^{27}{\rm Al})_{\rm disk} } \times
    \frac{ M({}^{27}{\rm Al})_{\rm disk}}{ M({}^{16}{\rm O})_{\rm disk}  }. 
 \label{eq:x}
\end{equation}
Most of these terms are defineable. 
First, $M({}^{26}{\rm Al})_{\rm SN} / M({}^{26}{\rm Al})_{\rm disk} \equiv \exp( +\Delta t / \tau)$,
where $\Delta t$ is the time delay between supernova injection and isotopic closure of the
meteoritic materials, and $\tau = 1.03 \, {\rm Myr}$ is the mean lifetime of $\al$.
By definition, $M({}^{26}{\rm Al})_{\rm disk} / M({}^{27}{\rm Al})_{\rm disk} \equiv$
$(26/27) \times (5 \times 10^{-5})$, because sufficient $\al$ must be injected to
yield the meteoritic ratio. 
Finally, the isotopic abundances in the solar nebula are known (the ratio 
${}^{27}{\rm Al} / {}^{16}{\rm O}$ is taken from \citet{Lodd03}), so 
% $\alratio \equiv 5 \times 10^{-5}$ as in MacPherson et al.\ 1995), 
we derive 
\begin{equation}
x = \frac{ 4.846 \times 10^{-7} }{ [ M({}^{26}{\rm Al}) / M({}^{16}{\rm O})]_{\rm SN} } \, \exp( +\Delta t / \tau).
\end{equation}
Note that $x$ is independent of the mass of the disk, but it increases with $\Delta t$, since 
larger values of $\Delta t$ imply that more supernova material had to be injected to yield the
same $\alratio$ ratio, thereby implying larger isotopic shifts in oxygen associated with this
injection.

Besides the time delay $\Delta t$, the major inputs needed to infer $(\dxvii,\dxviii)_{0}$ are
the isotopic composition $(\dxvii,\dxviii)_{\rm SN}$ and ratio of ${}^{16}{\rm O}$ to ${}^{26}{\rm Al}$ 
in the supernova ejecta, and the oxygen isotopic composition of the post-injection solar nebula.
GM07 used bulk abundances of supernova ejecta calculated by \citet{Raus02}
for the first set of quantities.
They also assumed that the oxygen isotopic ratios of the post-injection solar nebula matched the SMOW 
values of the present-day Earth: $(\dxvii',\dxviii') = (0\, \promille, 0\, \promille)$. 
This assumption is the main reason why they concluded that the pre-injection solar nebula had to be 
${}^{17}{\rm O}$-rich, as we now demonstrate. 
Rearranging equation~\ref{eq:shift} yields
\begin{equation}
\delta^{17}{\rm O}_{0} = \delta^{17}{\rm O}' 
 +x \left( \delta^{17}{\rm O}' - \delta^{17}{\rm O}_{\rm SN} \right). 
\end{equation} 
Supernova ejecta tend to be ${}^{16}{\rm O}$-rich; in the extreme limit, 
$\delta^{17}{\rm O}_{\rm SN} \approx -1000\, \promille$.  
If $\delta^{17}{\rm O}' \approx 0\, \promille$ also, then 
$\delta^{17}{\rm O}_{0} \approx +(1000 x)\, \promille$.
That is, $\delta^{17}{\rm O}_{0}$ is inferred to have been positive and potentially
quite large if $x > 10^{-2}$. 
The isotopic yields of the supernova ejecta computed by \citet{Raus02}
were consistent with such large values of $x$ and $\delta^{17}{\rm O}_{\rm SN} < 0$, 
leading GM07 to conclude that generally $\delta^{17}{\rm O}_{0} > 0\, \promille$.
Indeed, for progenitor masses $15 - 25 \, M_{\odot}$, GM07 inferred
$\delta^{17}{\rm O}_{0} \approx +35$ to $+220\, \promille$.
Since there are no early-formed meteoritic components with $\delta^{17}{\rm O}$ this
high, and because the oxygen isotopic composition of the Sun appears to be consistent
with $\delta^{17}{\rm O} \approx -60\, \promille$ \citep{Mcke09}, GM07 ruled
out supernova injection of $\al$ and $\ca$.
This conclusion depends on a few key assumptions that we update below.
We consider the starting composition of the solar nebula, and take into account the 
non-homogeneity of supernova ejecta. 

\subsection{Solar Nebula Oxygen Isotopic Composition}

Oxygen isotopic ratios potentially can test or rule out the supernova injection 
hypothesis, but several caveats must be applied to the method of GM07.
The first and most important correction involves the oxygen isotopic composition of the 
solar nebula immediately before and after the injection of supernova material.
GM07 assumed the post-injection composition was equal to SMOW; 
however, SMOW is widely understood {\it not} to reflect the oxygen 
isotopic ratios of the solar nebula immediately after injection.
On a three-isotope diagram of $\dxvii$ versus $\dxviii$, the oxygen isotopes of planetary and
meteoritic materials are arrayed along a mixing line called the Carbonaceous Chondrite Anhydrous
Mineral (CCAM) line discovered by \citet{Clay73}.
After correcting for isotopic fractionation by thermal and chemical processes, \citet{YounR98}
inferred a mixing line with slope 1.0 in the three-isotope diagram, and so we will
refer to this mixing line as the ``slope-1" line.
Today the oxygen isotopic composition of the Earth (SMOW) is widely recognized to reflect a mixture
of an isotopically lighter rocky component (to which CAIs belong), and an isotopically heavy reservoir
\citep[e.g.][]{Clay03,YounR98}.
It is very likely that this component is isotopically heavy water, with
$\dxvii,\dxviii > +30\, \promille$
\citep{Clay84,LyonY05,Lyonea09}.
The existence of isotopically heavy water is supported by the discovery (in the primitive
carbonaceous chondrite Acfer 094) of a poorly characterized product of aqueous alteration, with
$\dxvii \approx \dxviii \approx +180\, \promille$ \citep{SakaS07}.
Quite possibly this heavy water is the result of a mass-dependent photodissociation of CO in the outer
solar nebula by an external ultraviolet source \citep{LyonY05,Lyonea09}.
The photodissociation can be isotopically selective because the different isotopologues of CO
molecules can self-shield; ${\rm C}^{17}{\rm O}$ and ${\rm C}^{18}{\rm O}$ are optically thin and 
dissociate more completely, releasing ${}^{17}{\rm O}$ and ${}^{18}{\rm O}$ atoms that react with 
${\rm H}_{2}$ to form isotopically heavy water, while the abundant molecule ${\rm C}^{16}{\rm O}$ 
is more optically thick and does not as completely dissociate.
The light CO molecule is eventually lost with the nebular gas.
Whatever the source of the isotopically heavy component, 
SMOW only represents a late stage in nebular evolution, and does not represent
the state of the nebula immediately after injection of supernova material. 

Applying the same reasoning, it is likely that the starting composition of the solar nebula
was lower (more ${}^{16}{\rm O}$-enriched) on the slope-1 line than most CAIs. 
The majority of CAIs tend to cluster near $(\dxvii,\dxviii) \approx$ 
$(-41\, \promille,-40\, \promille)$, i.e., $\Delta^{17}{\rm O} \approx -20.2\, \promille$
\citep[see][and references therein]{Clay03};
but many of the most primitive and unaltered CAIs cluster near 
$(\dxvii,\oxviii) \approx$ $(-50\, \promille,-50\, \promille)$,
or $\Delta\oxvii \approx -24\, \promille$ \citep{ScotK01}.
Likewise, \citet{MakiN09} report $\Delta\oxvii = -23.3 \pm 1.9\, \promille$ for
``mineralogically pristine" CAIs. 
CAIs also contain grains of hibonite, spinel, and corundum, which are among the first 
minerals expected to condense from a cooling gas of solar composition \citep{EbelG00},
and which are presumably even more primitive than CAIs themselves. 
\citet{ScotK01} report that hibonite grains are also found to cluster near
$(\dxvii,\dxviii) \approx (-50\, \promille,-50\, \promille)$,
or $\Delta^{17}{\rm O} = -24 \,\promille$, while \citet{MakiN09b}
observed 4 hibonite grains from Allende and Semarkona to have oxygen isotopes
in the range $\Delta^{17}{\rm O} = -32$ to $-17\, \promille$. 
They also found that spinel grains from the CV chondrite Allende had 
$\Delta^{17}{\rm O} = -25 \pm 5\,\promille$, and that corundum grains from the CM chondrite 
Semarkona clustered strongly in the range $\Delta^{17}{\rm O} = -24 \pm 2\, \promille$. 
\citet{KrotN10} likewise report $\Delta^{17}{\rm O} = -24 \pm 2\, \promille$ for primitive CAIs 
and amoeboid olivine aggregates, which are also believed to have condensed from solar nebula gas. 
From these results we infer that $(\dxvii',\dxviii') \approx (-50\, \promille,-50\, \promille)$
in the solar nebula immediately after the injection of supernova material. 

Meteoritic and other samples also constrain the initial (pre-injection) oxygen isotopic 
composition of the solar nebula, and find it to be very similar. 
As described above, very firm and low upper limits to initial $\alratio$ exist for
FUN CAIs that mark them as having formed before the injection of $\al$ and $\ca$
\citep{SahiG98}. 
\citet{KrotC08} have identified a fractionation line associated with the FUN CAIs
with $\Delta^{17}{\rm O} = -24.1\, \promille$ that passes through
$(\dxvii,\dxviii) \approx (-51\, \promille,-52\, \promille)$.
Presumably the original isotopic composition of the nebula matched that of the Sun,
which might therefore be measured by {\it Genesis} mission \citep{Burn03}.
Preliminary measurements can be interpreted as clustering on a fractionation line
with $\Delta^{17}{\rm O} \approx = -26.5 \pm 5.6\, \promille$ \citep{Mcke09}, 
which would intersect the slope-1 line at 
$(\dxvii,\dxviii) \approx (-56\, \promille, -57\, \promille)$, and other analyses
suggest $\Delta^{17}{\rm O} \approx -33 \pm 8 \, \promille$ [$2\sigma$ errors]
\citep{McKeea10}. 

From these results it seems likely that the original solar nebula oxygen isotopic
composition was near the {\it Genesis} preliminary result of
$(\dxvii,\dxviii) \approx (-54\, \promille,-53\, \promille)$, or possibly much lower
along the slope-1 line. 
Subsequent reaction of rock with a ${}^{16}{\rm O}$-depleted reservoir 
then moved material along the slope-1 line to $(\dxvii,\dxviii) \approx$ 
$(-41\, \promille,-40\, \promille)$,
where most CAIs are found \citep[][and references therein]{Clay03}.
FUN CAIs appear to represent an intermediate stage in this process, only partially 
evolved along the slope-1 line. 
To fix values, we will simply assume the solar system protoplanetary disk isotopic
ratios started as $(\delta\oxvii,\delta\oxviii) = (-60\,\promille,-60\,\promille)$.

The above discussion changes the criterion by which one can reject the supernova
injection hypothesis. 
Because GM07 assumed an initial solar nebula composition near SMOW, 
they concluded that pre-injection samples necessarily would have had $\dxvii > 0$, 
and the lack of such samples in meteorites ruled out the hypothesis.
But we assert that the supernova injection hypothesis can be ruled out only if 
injection of supernova material necessarily shift the oxygen isotopic composition of 
the solar nebula from a composition near $(\dxvii,\dxviii) \approx (-60\, \promille,-60\, \promille)$ 
to one far off the slope-1 line, or one on the slope-1 line but with $\dxvii > -50 \, \promille$.
In this way, the GM07 method of using oxygen isotopic constraints might still allow
a test of the supernova injection hypothesis. 

\subsection{Magnitude of Isotopic Shift \label{magofshifts}}

There are at least three scenarios wherein the shift in oxygen isotopes following injection of 
supernova material can be consistent with the above constraints.
From equation~\ref{eq:shift}, it is seen that even if the supernova ejecta and the protoplanetary 
disk differ in oxygen isotopic composition by hundreds of permil, the shift in oxygen isotopes 
may be small ($< 1$ permil) if the injected mass is small, so that $x < 10^{-2}$. 
More precisely, if $\al$ and the other SLRs are injected by a supernova into the solar nebula
disk, then the magnitude of the shift in oxygen isotopes will depend on the fraction of 
ejecta oxygen that accompanies Al. 
In this first scenario, O and Al may be significantly fractionated during delivery of the ejecta 
to the solar nebula.
For example, \citet{Ouel07} find that effectively only material condensed from the supernova 
ejecta into large ($> 1 \, \mu{\rm m}$ radius)  grains can be injected directly into a protoplanetary 
disk.
In the extreme event that the only grains that entered the protoplanetary disk were corundum
(${\rm Al}_{2}{\rm O}_{3}$) grains, the isotopic shifts in oxygen would be negligible
($< 0.001\,\promille$).
Or, if only 10\% of the oxygen in the ejecta condensed into grains, and 90\% remained in gas
that was excluded from the disk, then the isotopic shifts in oxygen isotope (for a given amount
of injected $\al$) would be 10 times smaller than predicted by GM07. 
It is therefore not possible to determine the shifts in oxygen isotopes following injection into 
a disk without quantifying the degree to which O and Al are fractionated between gas and solids. 
In what follows, we assume no fractionation, as such a calculation is beyond the scope of the 
present investigation; but we consider dust condensation in supernova ejecta to be a very important 
effect, one that potentially could significantly reduce the predicted isotopic shifts. 

In the second scenario, the shifts in oxygen isotopes could also remain small if the injected 
material was simply higher than expected in $\al$ (or lower in O), so that again $x < 10^{-2}$. 
The calculations of GM07 relied on the {\it bulk} abundances calculated by \citet{Raus02}.
That is, GM07 assumed that the injected material uniformly sampled the entirety of the supernova
ejecta.
Such a uniform sampling is unlikely, as supernovae often do explode in 
a clumpy fashion and asymmetrically. It has long been understood that asymmetries or hydrodynamic 
instabilities may disrupt the stratification of the progenitor star, but they do not result in large scale 
compositional mixing \citep[e.g.][]{joggerst08, hftm05, fam91}.
The X-ray elemental maps of the Cassiopeia A supernova remnant \citep{Hwan04} dramatically 
demonstrate that massive stars are likely to explode as thousands of clumps of material, each sampling
different burning zones within the progenitor.
\citet{OuelDH10} have argued that this may be a near-universal feature of core-collapse
supernovae; at the very least, observations do not rule out this possibility. 
So it is more than possible that the solar nebula received materials from only limited regions 
within the ejecta in which the $\al / {}^{16}{\rm O}$ ratio could have varied considerably from
the average value for the ejecta. 
The non-uniformity of the $\al / {}^{16}{\rm O}$ ratio may be magnified if the star explodes
asymmetrically, allowing explosive nucleosynthesis to proceed differently even in parcels of 
gas in the same burning zone. 

Finally, in the third scenario by which isotopic shifts may conform to measurements, $x$ need not
be small, and the isotopic shifts may approach $10\, \promille$ in magnitude,
so long as the injection moved the composition {\it up} the slope-1 line by $\approx 10\, \promille$ 
(i.e., the change in $\dxvii$ equalled the change in $\dxviii$, both being $< 10\, \promille$), 
or {\it down} the slope-1 line by a comparable or even larger amount.
A shift from an initial composition $(\dxvii,\dxviii)_{0} \approx (-60\, \promille, -60\, \promille)$,
consistent with ${\it Genesis}$ measurements of the Sun's composition, to 
$(\dxvii,\dxviii)' \approx (-50\, \promille, -50\, \promille)$, consistent with primitive meteoritic
components, would not conflict with the data.
Alternatively,  a shift from an initial composition $(\dxvii,\dxviii)_{0} \approx (-60\, \promille, -60\, \promille)$,
to $(\dxvii,\dxviii)' \approx (-70\, \promille, -70\, \promille)$, 
or even $(\dxvii,\dxviii)' \approx (-80\, \promille, -80\, \promille)$, 
followed by mixing with the ${}^{16}{\rm O}$-poor reservoir that moves solar nebula solids up the 
slope-1 line, would also conform to the data.  

In the next section we compute the isotopic yields in core-collapse supernovae of various progenitor
masses, both in spherically symmetric explosions \citep[as considered by][]{Raus02}
and asymmetric explosions.
These calculations allow us to predict the oxygen isotopic composition $(\dxvii,\dxviii)_{\rm SN}$ 
and the ratio $x$ of the supernova material at various locations within the explosion, to assess the 
range of possible isotopic shifts under the second and third scenarios. A supernova injection scenario
would be ruled out, {\it unless} either the injection of material results in small overall shifts (i.e. the 
injected material contains a high \al~ abundance relative to oxygen, or vice versa, a low oxygen 
abundance relative to Al), or the oxygen isotopes are shifted along the 
'slope-1 line', in which case shifts of up to $\sim$10 permil in either direction are allowed.

%=========================================================================================%
% SECTION 3: ISOTOPE PRODUCTION WITHIN SUPERNOVAE                                         %
%=========================================================================================%

\section{Isotope Production within Supernovae}

%==========================================================================================
% TABLE 1
\begin{deluxetable}{lcccc}
\tablewidth{0pt}
\tablecaption{Explosion Simulations\label{tab1}}
\tablehead{
  \colhead{Simulation}
& \colhead{progenitor}
& \colhead{Energy}
& \colhead{M$_{\rm Remn}$}
& \colhead{Delay} \\
  \colhead{}
& \colhead{}
& \colhead{$10^{51}$\,erg}
& \colhead{(\Msol)}
& \colhead{ms}
}
\startdata
23e-0.8 & 23 \Msol\, single & 0.8 & 5.7 & 20 \\
23e-1.2 & 23 \Msol\, single & 1.2 & 4.1& 20 \\
23e-1.5 & 23 \Msol\, single & 1.5 & 3.2 & 20 \\
23e-0.7-0.8 & 23 \Msol\, single & 0.8 & 3.2 & 700 \\
23e-0.7-1.5 & 23 \Msol\, single & 1.5 & 2.3 & 700 \\
16m-run1 & 16 \Msol\, binary &  1.5 &  2.06 & 20 \\
16m-run2 & 16 \Msol\, binary &  0.8 &  2.43 & 20 \\
16m-run4 & 16 \Msol\, binary &  5.9 &  1.53 & 20 \\
23m-run1 & 23 \Msol\, binary &  1.9 &  3.57 & 20 \\
23m-run2 & 23 \Msol\, binary &  1.2 &  4.03 & 20 \\
23m-run5 & 23 \Msol\, binary &  6.6 &  1.73 & 20 \\
40m-run1 & 40 \Msol\, single &  10 &  1.75 & 20 \\
40m-run5 & 40 \Msol\, single &  1.8 &  4.51 & 20 \\
40m-run9 & 40 \Msol\, single &  2.4 &  6.02 & \,20
\enddata
\label{tb:explsim}
\end{deluxetable}
%==========================================================================================

\subsection{Numerical Methods}

We calculate the yields of oxygen isotopes, \ca, and \al\, in several core-collapse
supernova scenarios, listed in Table \ref{tb:explsim}.
These calculations explore four different progenitor models, with a range of explosion scenarios
for each.
We use a large set of thermally driven 1D explosions with varying kinetic energies and delays,
for a star of initial mass 23 $\Msol$ and a more restricted range of explosions for a 16 $\Msol$, 
and 23 $\Msol$, with the hydrogen envelope stripped in a case B binary scenario, and a single 40 
$\Msol$ progenitor that ends its life as a type WC/O after extensive mass loss.
We also examine a 3D explosion of the 23 $\Msol$ binary progenitor.
Details of the simulations can be found in \citet{yca09}.
The set of progenitor models we selected by no means samples the entire diversity of supernovae, but it
represents a variety of cases across a large range of progenitor masses and explosion parameters.
It is sufficiently diverse to make generalizations for behaviors that appear across all models.

Progenitor models were produced with the TYCHO stellar evolution code
\citep{YouA05}.
To model collapse and explosion, we use a 1-dimensional Lagrangian code developed by \citet{her94}
to follow the collapse through core bounce.
This code includes 3-flavor neutrino transport using a flux-limited diffusion calculation and a
coupled set of equations of state to model the wide range of densities in the collapse phase
\citep[see][for details]{her94,fryer99a}.
It includes a 14-element nuclear network \citep{bth89}
to follow the energy generation.
To get a range of explosion energies, we opted to remove the neutron star and drive an explosion by
injecting energy in the innermost 15 zones (roughly 0.035 $\Msol$).
The duration and magnitude of energy injection of these artificial explosions were altered to produce the
different explosion energies.
Our 3-dimensional simulation uses the output of the 1-dimensional explosion (23 $\Msol$ binary star, 23m-run5)
when the shock has reached 10$^9$\,cm.
We then map the structure of this explosion into our 3D Smooth Particle Hydrodynamics code SNSPH
\citep{frw06}
by placing shells of particles whose properties are determined by
the structure of our 1-dimensional explosion \citep[see][for details]{hun05, frw06}.
The 3D simulation uses 1 million SPH particles and is followed for 800 seconds after collapse.

The initial intent of the 3D simulation was to create a fully 3-dimensional
calculation of an explosion with a moderate bipolar asymmetry \citep{Younea06}.
The interesting behavior of \al\, in the explosion then prompted us to 
consider the composition of this material in relation to an isotopic enrichment scenario of
the solar system.
Both observational and theoretical evidence indicate that asymmetry is strong and ubiquitous in supernovae 
\citep[e.g.][]{grb07,yf07,hun05, Lopeea09}. In the situation of supernova injection the asymmetric 
model's primary utility lies not in modeling a specific event, but rather sampling a wide range of 
thermodynamic histories for material capable of producing \al. Injected material is likely to sample 
only a small region of the supernova, meaning we can treat each SPH particle as an isolated trajectory 
whose further evolution is not dependent upon the progenitor star or or global parameters of the 
explosion. Any explosion that produces a similar thermodynamic trajectory will end up with similar 
yields. We can thus probe a large variety of explosion condition not accessible to 1D calculations 
without a prohibitive investment of computational time. Therefore we consider this asymmetric
simulation sufficiently generic to justify its usage for this investigation.
We create an asymmetric explosion with a geometric aspect ratio and final kinetic energy axis ratio 
designed to be roughly consistent with the degree of asymmetry implied by supernova polarization 
measurements. To simulate an asymmetric explosion, we modify the velocities within each shell by 
increasing those of
particles within 30$^\circ$ of the z-axis by a factor of 6; we will refer to
these parts of the supernova as high velocity structures (HVSs). The velocities of the remaining particles were
decreased by a factor of 1.2, roughly conserving the explosion energy. This results in a 2:1 morphology
between the semimajor to semiminor axes ratio by the end of the simulation.
We did not introduce any angular dependence in the thermal energy.
At these early times in the explosion, much of the explosion energy remains in thermal energy, so the total
asymmetry in the explosion is not as extreme as our velocity modifications suggest.
For a detailed discussion on choosing asymmetry parameters for 3D explosions see
\citet{hun03,hftm05,fw04}.

%~~~~~~~~~~~~~~~~~~~~~~~~~~~~~~Figure~~~~~~~~~~~~~~~~~~~~~~~
\begin{figure}[tbh]
\centering      
\subfloat[]{\label{fig:dens}\includegraphics[angle=90,width=0.49\columnwidth]{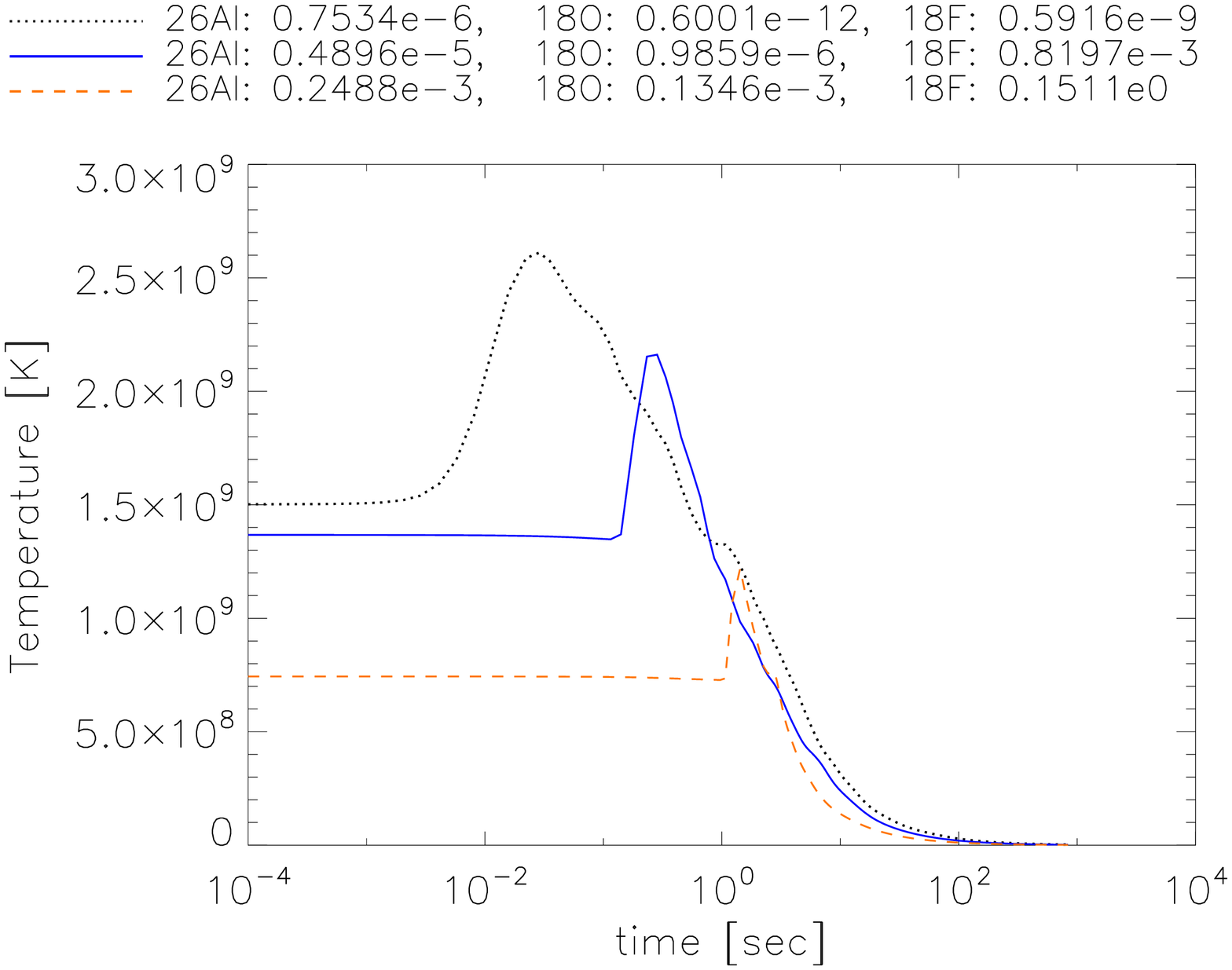}}\\

\subfloat[]{\label{fig:temp}\includegraphics[angle=90,width=0.49\columnwidth]{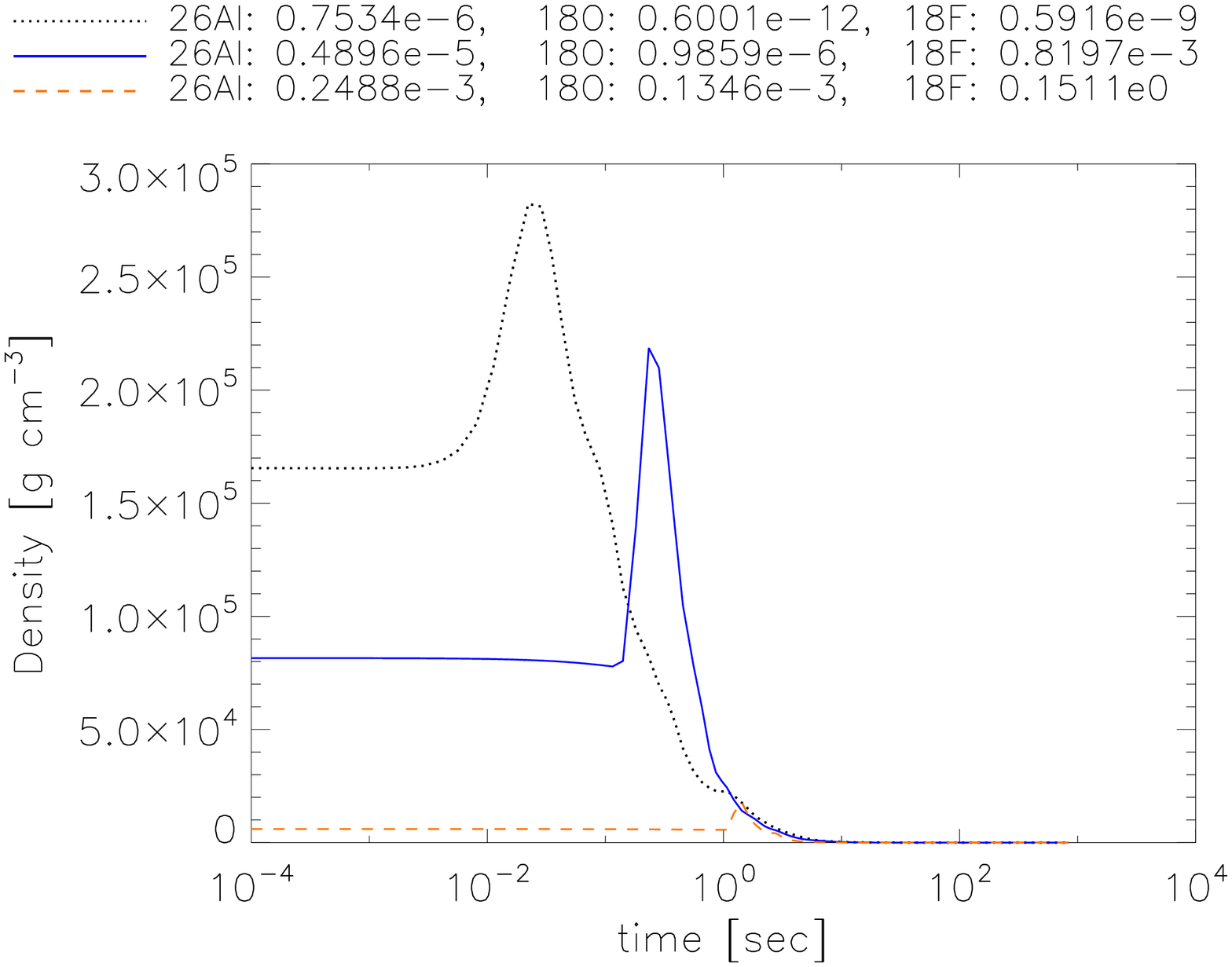}}\\

\subfloat[]{\label{fig:entr}\includegraphics[angle=90,width=0.49\columnwidth]{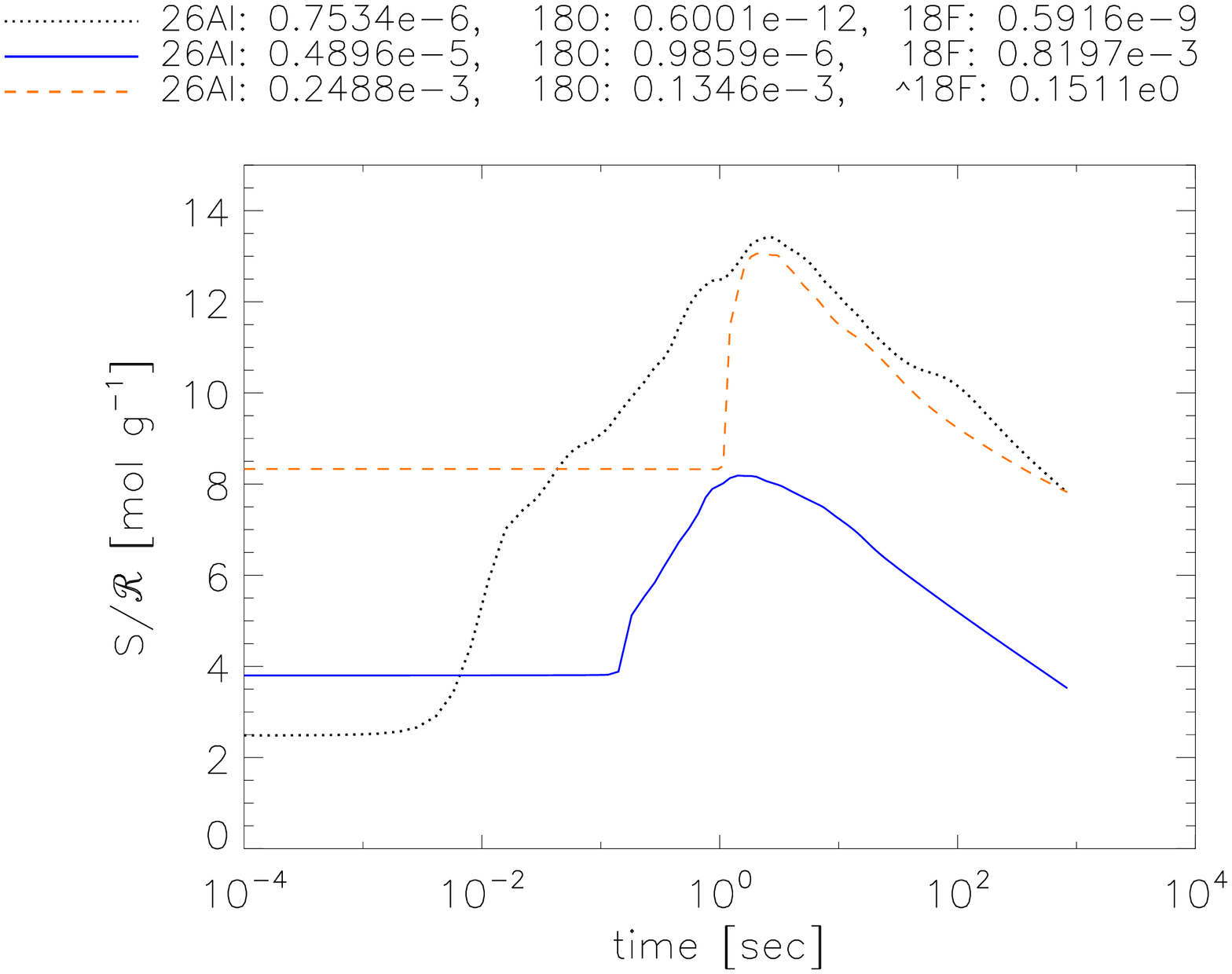}}
\caption{Shown is the time evolution of temperature, density, and entropy 
for representative examples of particles from
the Ring and Bubble region of the 3D calculation. Each line is labelled with the abundance of 
\al, \oxviii, and \fxviii~(above each graph) in mass fraction per particle at the end of 
the simulation.}
\label{fig:traj}
\end{figure}
%~~~~~~~~~~~~~~~~~~~~~~~~~~~~End~Figure~~~~~~~~~~~~~~~~~~~~~

As noted above, although we are only considering one 3D model, the results we get from that 
model are representative of a range of nucleosynthetic conditions that may occur in multiple 
explosion/progenitor scenarios. Each parcel of gas follows its 
own density and temperature evolution, which is determined by the \emph{local} velocity of the 
parcel of gas. It is the local conditions of the gas that matter; it is unaware of the global evolution.  
An example of the trajectories from the Ring and Bubble regions are shown in Figure 
\ref{fig:traj}. The Figure shows the temperature, density, and radiation entropy evolution for 
representative $\al$-rich particles in the explosion. The lines are labeled with $\al$, $^{18}$F, 
and $\oxviii$ abundances at the emd of the simulation (before complete radioactive decay of $^{18}$F). 
We see three classes of trajectories: high temperature and high entropy, high temperature and 
low entropy, and low temperature, high entropy. Predictably, the high temperature, low entropy 
trajectories tend to have low $^{18}$F (and therefore $\oxviii$) abundances due to the 
photodisintegration of $^{18}$F into $^{14}$N $+\ \alpha$. High temperature, high entropy 
trajectories have a higher reverse rate for that reaction, preserving slightly more $^{18}$F. The low 
temperature particles have the highest $^{18}$F abundance at the end of burning.

The asymmetric explosion samples trajectories with a large span 
of velocity evolutions reasonable for plausible asymmetries. As we demonstrate with the 1D models and in 
\citet{yca09}, the sites of production for \al~ are similar across a wide range of stellar 
masses, so as long as we sample the particle trajectories well in a single asymmetric explosion, the 
results are robust to very large changes in progenitor mass and explosion asymmetry. 
This assumption is valid, since we are 
looking for regions in the explosion that produce plausible abundances, not a bulk yield.

The network in the explosion code terminates at $^{56}$Ni and cannot follow neutron excess, so to
accurately calculate the yields from these models we turn to a post-process step.
Nucleosynthesis post-processing was performed with the Burn code \citep{yf07},
using a 524 element network terminating at $^{99}$Tc.
The initial abundances in each SPH particle are the 177 nuclei in the initial stellar model.
The network machinery is identical to that in TYCHO \citep[for details of the simulations see][]{yca09}.

%~~~~~~~~~~~~~~~~~~~~~~~~~~~~~~Figure~~~~~~~~~~~~~~~~~~~~~~~
\begin{figure}[ptbh]
\centering
\includegraphics[angle=0,width=0.69\columnwidth]{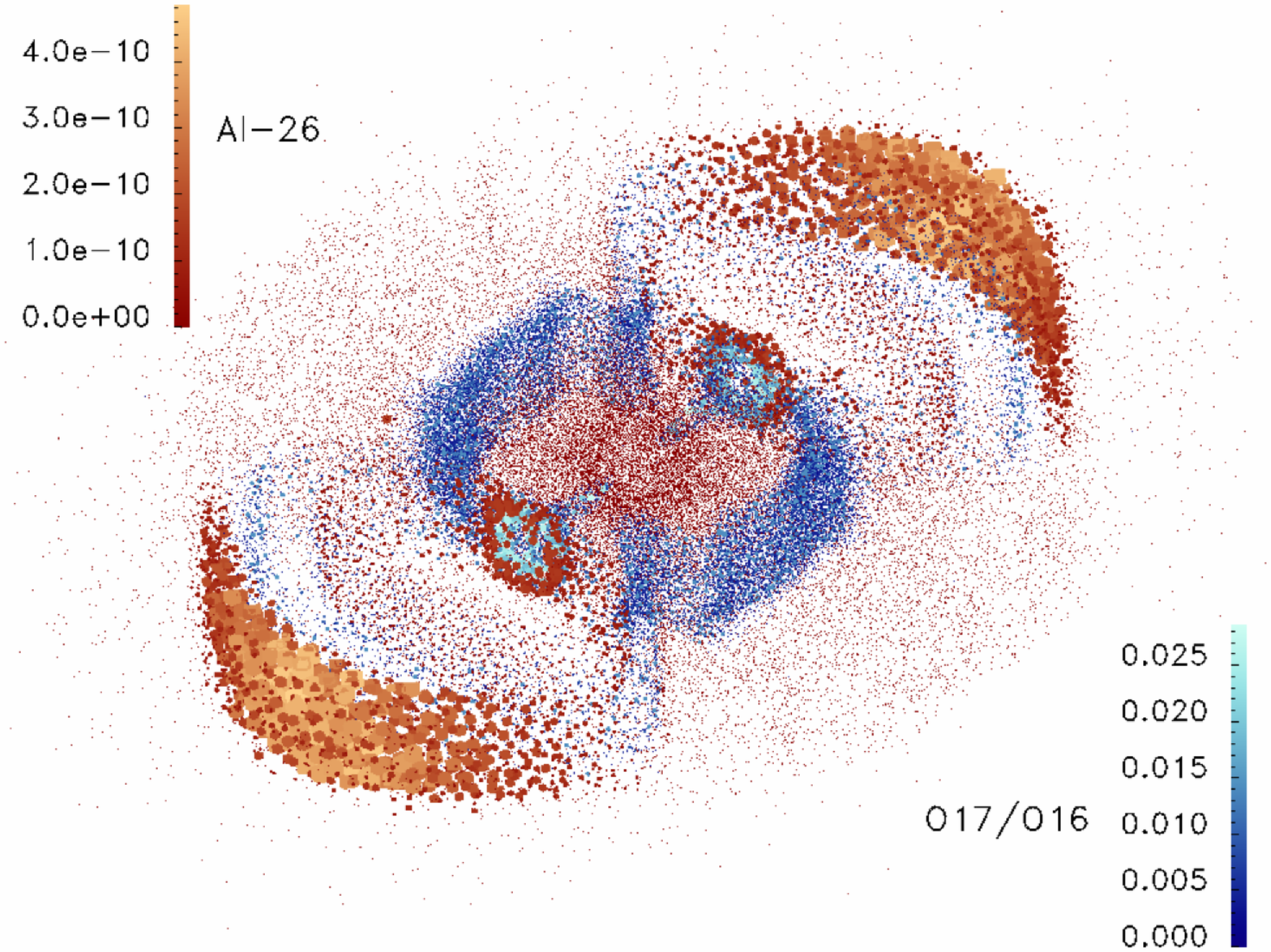}
\\
\includegraphics[angle=0,width=0.69\columnwidth]{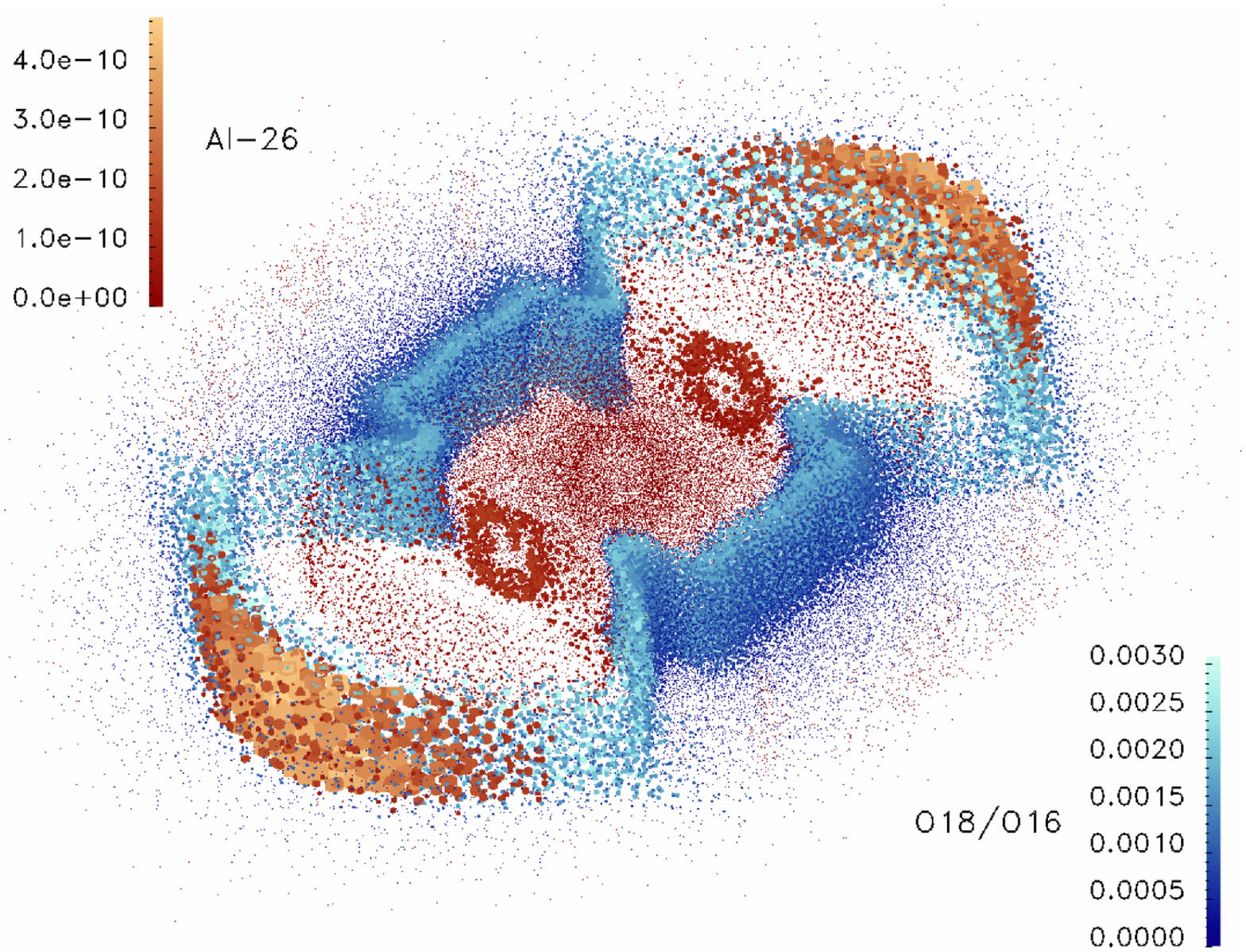}
\caption{Shown is a $2.0\times10^{11}$ cm thick slice in the x-z plane of
the 3D simulation. \al\, abundances are in red-tones (amount per particle in \Msol) and number
ratio of $^{17}$O/$^{16}$O (top panel) or $^{18}$O/$^{16}$O (bottom panel)
per particle in blue-tones. The sizes of the data points are arbitrarily chosen, but
scale with their values. A color gradient was also used to visualize the different
abundances per particle. The ligher colors/bigger data points correspond to
higher abundances.
Apparent is a Ring on either side of the center along the axis of symmetry,
and further out from them the Bubbles, where the highest \al\, abundance is found.}
\label{fig:al}
\end{figure}
%~~~~~~~~~~~~~~~~~~~~~~~~~~~~End~Figure~~~~~~~~~~~~~~~~~~~~~

%~~~~~~~~~~~~~~~~~~~~~~~~~~~~~~Figure~~~~~~~~~~~~~~~~~~~~~~~
\begin{figure}[tbh]
\centering
\includegraphics[angle=0,width=0.7\columnwidth]{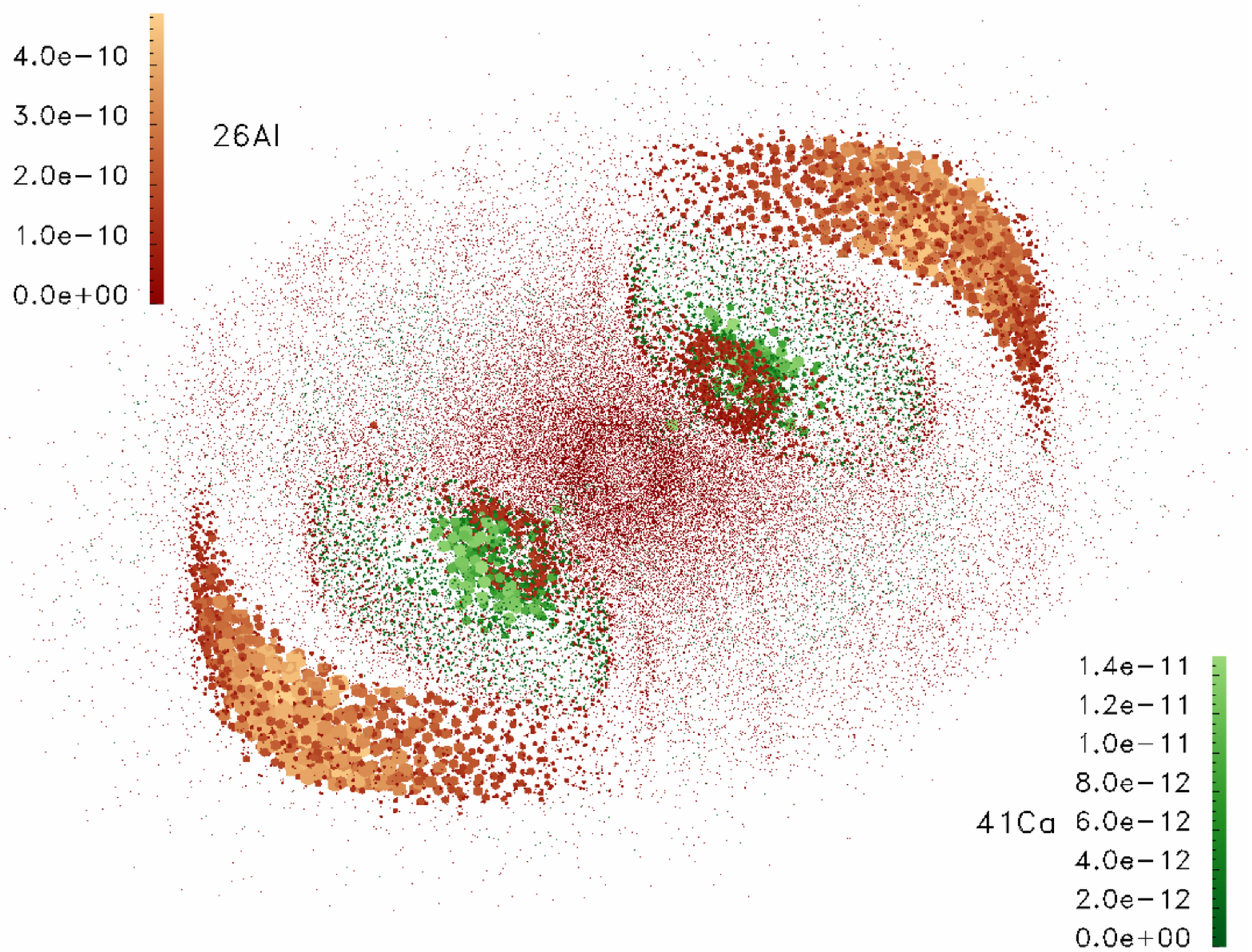}
\caption{Same as Figure \ref{fig:al} but with $^{41}$Ca shown in green. The highest \ca\, abundance
is adjacent to the Rings, and only partially overlaps with the Ring- regions.}
\label{fig:ca}
\end{figure}
%~~~~~~~~~~~~~~~~~~~~~~~~~~~~End~Figure~~~~~~~~~~~~~~~~~~~~~

\subsection{Production of ${}^{26}{\rm Al}$ in 1D and 3D explosions}

There are three primary sites for production of \al\, in a massive star and its accompanying
supernova.
It can be produced by hydrogen burning at high temperatures in the shell-burning regions of
massive stars or evolved AGB stars. But neither of these production sites
is important to the supernova injection scenario, as discussed in \citet{yca09}.
In the 1D simulations, the two dominant production sites are two peaks in $^{26}{\rm Al}$
abundances that coincide with peak temperatures
in the explosion of  $2.2 \times 10^9 \, {\rm K}$ and $1.5 \times 10^9 \, {\rm K}$, in material
that has undergone hydrostatic C burning in the progenitor.
The higher of the two temperatures is sufficient for explosive C and Ne burning, the lower of the
temperatures is near explosive C burning.
The production of \al\, in both regions is due to a significant increase in the
flux of free p, n, and $\alpha$-particles.
At higher temperatures, characteristic of O burning, $^{26}{\rm Al}$ is quickly destroyed.

Within the 3D calculations, $^{26}{\rm Al}$ is produced in two main regions
(see Figure \ref{fig:al}), similar to the 1D results.
The first is a ring-like structure and an associated small bubble where the two
HVSs emerge into a lower density region, and which we denote the ``Ring".
Material in the Ring has undergone explosive Ne and C burning during the explosion at
temperatures slightly above $2 \times 10^9 \, {\rm K}$, and
corresponds to the explosive C and Ne region in the 1D simulations identified above.
The second region, further out at the terminal end of the HVSs, is denoted the ``Bubble"
(see Figure \ref{fig:al}).
Within the Bubble, material has undergone hydrostatic C burning and then experienced
peak shock temperatures $\sim 1.5 \times 10^9 \, {\rm K}$ during the explosion,
and corresponds to the second of the \al\, peaks in the 1D
simulations identified above (which we will refer to as sub-explosive C burning region).
While the production sites of \al\, in the simulation in 3D occur in zones of about the same
temperatures as in the 1D cases, the peak in \al\, abundance in those regions are reversed
from the corresponding regions in 1D (i.e. the peak that is higher in \al\, in 1D is lower in \al\,
in 3D, and vice versa). An important aspect of the Bubble is that due to the rapid expansion
of the HVSs, its density drops rapidly, quenching some of the nuclear reactions.
The decrease in density in the 1D simulations occurred at
a slower rate, conversely more of the \al\, was able to be processed into other species.
This freezeout of nuclear reactions (suppressing subsequent destruction of \al)
in the  3D simulation is the reason for the higher production of $^{26}{\rm Al}$
in the Bubble, as compared to the Ring.

In 3D, the \ca\, production occurs in only one main production site, in a region adjacent to and
partly overlapping the Ring in the 3D simulation (see Figure \ref{fig:ca}), and in both the 
explosive C/Ne and sub-explosive C burning regions in the 1D calculations. The production of 
\ca~ requires a high $^{40}$Ca abundance as the seed 
nucleus, and the main production channel is p- and n-capture onto $^{40}$Ca. 
The 
slower drop in density and temperature in the 1D calculations tended to favor a low level production of 
\ca, which is why its abundance is slightly higher as compared to the 3D calculation. In the 3D 
calculation, \ca~ is produced in the Ring, but the faster expansion of the material there due to the 
velocity asymmetry shuts off the reactions faster than in the 1D models, and the final \ca~ abundance
is lower than in 1D. In the bubble region, the lower temperature and rapid density falloff preclude 
any significant \ca~ production. 

Within the zones where \al\, is produced, the M$({}^{26}{\rm Al})$/M(${}^{16}{\rm O})$
ratios can differ significantly from the bulk abundances.
For the 1D models, these ratios can vary by a factor of $\sim 1$ up to a factor of
$\sim 100$ between the
explosive C/Ne burning region and the sub-explosive C burning region, with a typical
variation of a factor of $\sim 2-3$.
For example, in model 23e-1.5 the
M(\al)/M(\oxvi) ratio varies from $(6.7-8.4)\times10^{-6}$, and
for model 16m-run2 varies from $1.6\times10^{-7}$ to $1.7\times10^{-5}$.
These are to be compared to the abundances in the bulk of the ejecta, which are
M(\al)/M(\oxvi) $\approx 6.7 \times 10^{-6}$ for model 23e-1.5,
$\approx 1.7 \times 10^{-6}$ for model 16m-run2, and varies between
$4.0 \times 10^{-7}$ to $1.9 \times 10^{-4}$
across all 1D explosions. In the 3D model, the ratios are
M(\al)/M(\oxvi) $\approx 1 - 4 \times 10^{-4}$ in the SPH particles
in the Bubble, $\approx 1 - 4 \times 10^{-5}$ in the SPH particles
in the Ring, and
$2.88 \times 10^{-5}$ for the bulk supernova abundances.
Thus we see that injection of material from the Bubble brings in an order of
magnitude less oxygen (essentially all ${}^{16}{\rm O}$) per \al\, atom than injection
of material from the supernova overall or from the Ring.

\subsection{Production of O isotopes in ${}^{26}$Al-producing regions}

The abundances and isotopic compositions of oxygen within a localized region of the
supernova can vary significantly from the bulk values, as their production is sensitive to
the density, temperature, and composition, and to their variations with time in that region.
In the 1D explosions, density falls off roughly as a power law \citep{arnett96}.
Because this maintains a high density in the region where
\al\, forms by explosive C and Ne burning,
\oxviii\, is effectively synthesized into heavier species.
\oxvii\, is also synthesized into heavier species but is also created by neutron
captures onto \oxvi.
The net effect is that both \oxvii\, and \oxviii\, are reduced relative to \oxvi, and
the \oxviii/\oxvii\, ratio is reduced.
In the 23e-1.5 model, their mass fractions in the \al\, rich zones never exceed
$\sim 10^{-5}$, and the isotopic composition in nearly all our 1D cases
approaches $(-1000\,\promille,-1000\,\promille)$, effectively pure \oxvi.

The details of oxygen isotopic abundances in the \al\, rich zones of the 3D explosion
differ.
The more rapid expansion of the material in the 3D calculation limits the processing of 
\oxvii~ and other isotopes into heavier species, so the yield of those is higher than in the 1D calculations.
As one of the main burning products of explosive Ne burning, $^{16}$O is
quite abundant in the Ring.
However, some of the free p and n produced during explosive burning capture onto
\oxvi, producing $^{17}$O, so the Ring (as Figure \ref{fig:al} shows) is quite enriched
in \oxvii\, relative to the rest of the explosion. 
In the Bubble, \oxvi\, is not produced explosively, and is mostly
left over from the progenitor.
The increased flux of free particles also burns some of the $^{16}$O there to $^{17}$O
and $^{18}$O.
The freezeout from the expansion limits the processing of
these isotopes into heavier species, so the 3D explosion is richer in these isotopes than
the 1D simulation.

Part of the reason for the large variation in ${}^{18}{\rm O}$ isotopic yields is that
most of it is produced by decay of $^{18}{\rm F}$ ($t_{1/2} = 110$ minutes), which was
co-produced with \oxvii\, and \oxviii. 
Thus it depends sensitively on how much \fxviii~ is 
present once nuclear burning shuts off, which in turn depends sensitively on the 
trajectories taken by the gas.
At low temperatures the classical decay reaction $^{18}{\rm F} \, \rightarrow \, {}^{18}{\rm O} + e^{+}$ 
completely dominates, but at high temperatures (above $\sim 10^{9} \, {\rm K}$), 
and low proton density, another decay channel opens up for
${}^{18}{\rm F}$, and it
can decay also via 
$^{18}{\rm F} \, \rightarrow \, {}^{14}{\rm N} + \alpha$ \citep{ga00}.
The branching ratio of these two reactions is very sensitive to temperature at around $1\times10^9$ K,
with higher temperatures overwhelmingly favoring the decay to  ${}^{14}{\rm N} + \alpha$.

The amount of \fxviii~ remaining at the end of burning is highly dependent on the time taken 
to drop below that temperature, and the density evolution, as high entropies favor the destruction over 
the synthesis. 
Because of the power law drop off, the density in the 1D calculations stayed higher for a longer period of time,
as compared to the 3D calculation, thus isotopes had a longer time window in which they could be processed
to higher species. The density of the 3D calculation dropped faster due to the 
increased velocities of particles to create the asymmetry, so the nuclear burning shut off earlier, 
and more isotopes like \oxvii, \oxviii, or \fxviii~ survived the nucleosynthesis of the explosion.
This results in a substantial variation in the $^{18}{\rm F}$ abundance
between the 1D and the 3D calculation, and that same effect (i.e. how quickly the density drops) 
is also responsible for the variation in abundance 
of particles in the Ring and the Bubble by the end of the 3D simulation.

In the 1D simulations the full decay of all \fxviii\, after the explosion was calculated in the reaction network.
The 3D simulation was terminated earlier in its evolution before complete decay of the \fxviii. As the
 temperature at that point in the explosion was well below $10^9$ K,
we assumed that any \fxviii\, still present would decay into \oxviii, as this is the only significant 
channel at these lower temperatures.

%=========================================================================================%
% SECTION 4: ISOTOPIC SHIFTS                                                              %
%=========================================================================================%

\section{Solar system oxygen isotopic shifts accompanying ${}^{26}{\rm Al}$ delivery}

%~~~~~~~~~~~~~~~~~~~~~~~~~~~~~~Figure~~~~~~~~~~~~~~~~~~~~~~~
\begin{figure}[tbh]
\centering
\includegraphics[angle=90,width=0.8\columnwidth]{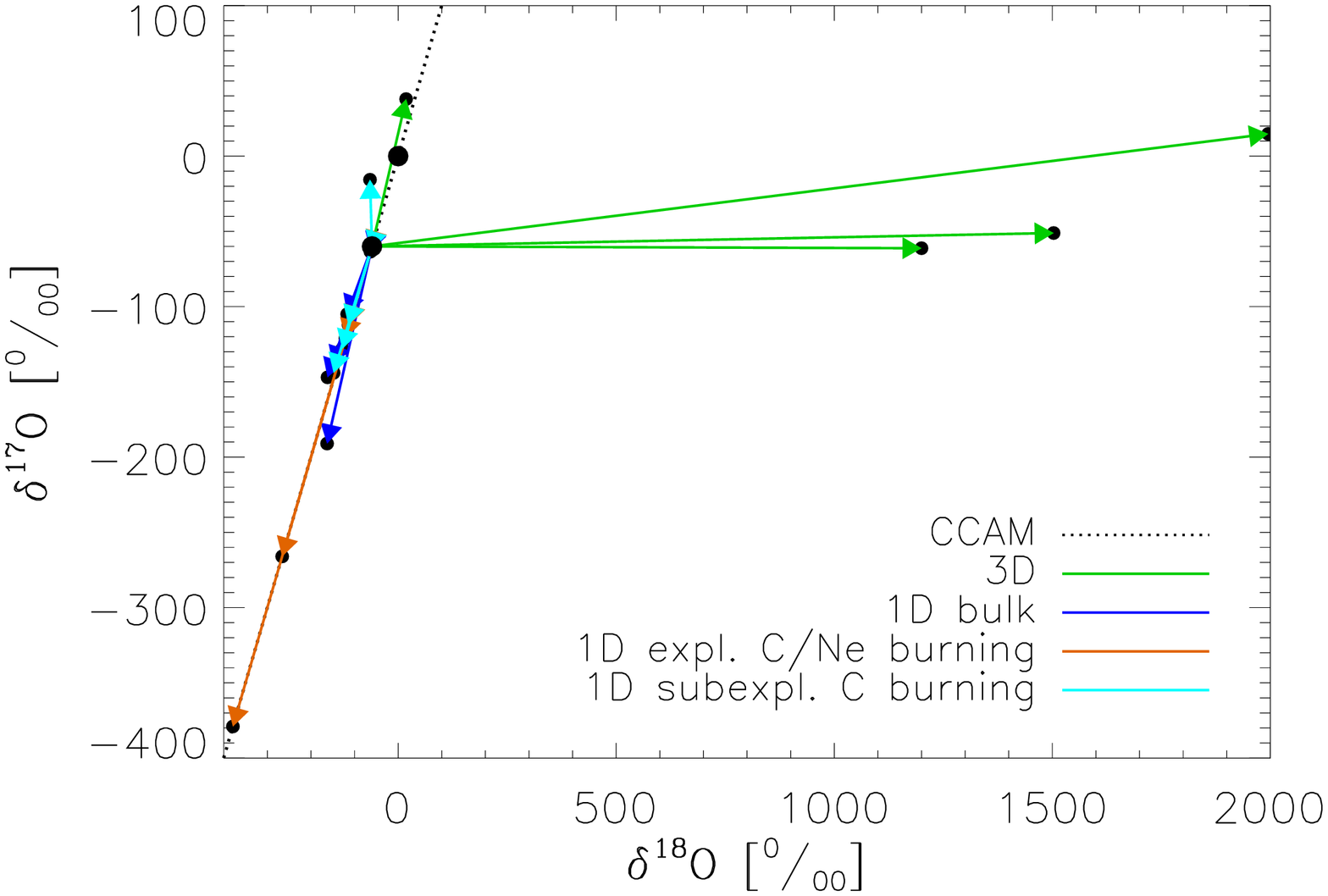}
\caption{Three isotope plot showing the shifts in the oxygen isotopes we calculate following
injection of supernova material for different scenarios. The shifts from he 3D 
cases are plotted in green, those from the 1D bulk cases are plotted in blue, the 1D
explosive C/Ne burning cases are plotted in orange, and the 1D sub-explosive C burning cases 
are plotted in cyan. Indicated by the bigger black dots are SMOW at 
($0\, \promille$, $0\, \promille$) and our assumed pre-injection composition at 
($-60\, \promille$, $-60\, \promille$). The very large shift of the 3D bulk scenario was omitted for clarity.}
\label{fig:3isob}
\end{figure}
%~~~~~~~~~~~~~~~~~~~~~~~~~~~~End~Figure~~~~~~~~~~~~~~~~~~~~~

\subsection{Spherically Symmetric Supernova Explosions}

It is now possible to calculate the shifts in oxygen isotopic abundances before and after
the injection of supernova material, using equation~\ref{eq:shift}.
As described in \S 2, the initial composition of the solar nebula was probably close to 
$(\dxvii,\dxviii)_{0} \approx (-60\, \promille,-60\, \promille)$, and we adopt this as 
the starting value.
Based on the numerical simulations of \S 3, we have calculated the ratios $x$ and the isotopic 
abundances $(\dxvii,\dxviii)_{\rm SN}$ within the ejecta overall, and within the regions where 
$\al$ is produced.

We begin with the case of the 1D explosions.
The regions where $\al$ is produced are those we identified as the C/Ne explosive burning region,
and the sub-explosive C burning region.
For our purposes, these regions were defined based on the $\al$ content. 
The exact amount of $\al$ produced varied among the simulations, but the (radial) abundance 
distribution of $\al$ in each simulation showed two distinct peaks that were at least one order 
of magnitude higher than the average $\al$ mass fraction. 
Thus the $\al$-rich regions in the 1D simulations were defined to be at least one order of 
magnitude higher in mass fraction than the average distribution.
The final isotopic composition of the solar nebula following injection of supernova material
has been calculated first assuming the material had the average (bulk) composition of the ejecta for comparison with \citet{Goun07},
and then that of one of these $\al$-rich regions.
The results are presented in Tables~\ref{tb:1Da}- \ref{tb:1Dc} and in Figure \ref{fig:3isob}.
Our results for injection of bulk ejecta from 1D explosions conforms closely to the findings 
of GM07 using the 1D models of \citet{Raus02}.
The ejecta are generally very ${}^{16}{\rm O}$-rich, with $\dxvii$ and $\dxviii$ that are 
large and negative.

%==========================================================================================
% TABLE 2 - 1D-individual
\begin{deluxetable}{lrrrrr}
 \tablewidth{0pt}
 \tablecaption{Oxygen isotopic shifts following injection from a 1D supernova}
 \tablehead{
 \colhead{$\,$}  & \colhead{16m-}    & \colhead{23m-}    & \colhead{40m-}    & \colhead{23e-}    & \colhead{23e-0.7-} \\
 \colhead{$\,$}  & \colhead{average} & \colhead{average} & \colhead{average} & \colhead{average} & \colhead{average} }
  \startdata
  \multicolumn{6}{c}{\underline{Bulk}}\\
  \\
  \oxvi & $1.97$ \Msol  & $1.48$ \Msol  & $3.29$ \Msol  & $2.93$ \Msol  & $5.44$ \Msol \\
  \oxvii        & $2.24\times10^{-4}$ \Msol     & $7.67\times10^{-6}$ \Msol     & $1.80\times10^{-4}$ \Msol     & $3.6\times10^{-4}$ \Msol      & $2.94\times10^{-4}$ \Msol \\
  \oxviii       & $1.85\times10^{-3}$ \Msol     & $1.53\times10^{-5}$ \Msol     & $3.29\times10^{-5}$ \Msol     & $6.63\times10^{-5}$ \Msol     & $1.27\times10^{-4}$ \Msol \\
  \al   & $3.86\times10^{-6}$ \Msol     & $1.91\times10^{-4}$ \Msol     & $1.52\times10^{-5}$ \Msol     & $2.17\times10^{-5}$ \Msol     & $2.15\times10^{-5}$ \Msol \\
  \ca   & $4.06\times10^{-6}$ \Msol & $2.63\times10^{-6}$ \Msol & $1.21\times10^{-5}$ \Msol & $1.06\times10^{-5}$ \Msol & $1.49\times10^{-4}$ \Msol \\
  \\
  \dxvii               & $-719\,\promille$ & $-987\,\promille$ & $-899\,\promille$ & $-788\,\promille$ & $-854\,\promille$ \\
  \dxviii              & $-581\,\promille$ & $-996\,\promille$ & $-997\,\promille$ & $-993\,\promille$ & $-990\,\promille$ \\
  $x$                  & 0.247           & 0.00375         & 0.105           & 0.0655          & 0.123           \\
  \\
  Final \dxvii         & $-191\,\promille$ & $-63.5\,\promille$ & $-139\,\promille$ & $-105\,\promille$  & $-147\,\promille$ \\
  Final \dxviii        & $-163\,\promille$ & $-63.5\,\promille$ & $-149\,\promille$ & $-117\,\promille$  & $-162\,\promille$ \\
  Final $\Delta\oxvii$ & $-118\,\promille$ & $-31.1\,\promille$ & $-65.6\,\promille$& $-45.1\,\promille$ & $-66.2\,\promille$ \\
  $\Delta$t 		& 1.41 Myr		& 0.65 Myr		& 1.36 Myr		& 1.31 Myr		& 1.73 Myr \\
\enddata
\label{tb:1Da}
\end{deluxetable}

\begin{deluxetable}{lrrrrr}
 \tablewidth{0pt}
 \tablecaption{Oxygen isotopic shifts following injection from a 1D supernova}
 \tablehead{
 \colhead{$\,$}  & \colhead{16m-}    & \colhead{23m-}    & \colhead{40m-}    & \colhead{23e-}    & \colhead{23e-0.7-} \\
 \colhead{$\,$}  & \colhead{average} & \colhead{average} & \colhead{average} & \colhead{average} & \colhead{average} }
  \startdata
  \multicolumn{6}{c}{\underline{explosive C/Ne burning}}\\
  \\
  \oxvi & $5.19\times10^{-1}$ \Msol     & $5.14\times10^{-1}$ \Msol     & $4.83\times10^{-1}$ \Msol     & $1.35$ \Msol  & $1.86$ \Msol \\
  \oxvii        & $1.47\times10^{-7}$ \Msol     & $2.89\times10^{-7}$ \Msol     & $2.76\times10^{-8}$ \Msol     & $8.79\times10^{-7}$ \Msol     & $4.96\times10^{-6}$ \Msol \\
  \oxviii       & $4.41\times10^{-7}$ \Msol     & $4.02\times10^{-6}$ \Msol     & $1.70\times10^{-8}$ \Msol     & $1.64\times10^{-8}$ \Msol     & $3.21\times10^{-8}$ \Msol \\
  \al   & $8.98\times10^{-7}$ \Msol     & $1.04\times10^{-4}$ \Msol     & $4.55\times10^{-7}$ \Msol     & $9.37\times10^{-6}$ \Msol     & $1.53\times10^{-5}$ \Msol \\
  \ca   & $1.74\times10^{-6}$ \Msol & $2.74\times10^{-6}$ \Msol & $5.59\times10^{-8}$ \Msol & $6.70\times10^{-6}$ \Msol & $1.16\times10^{-4}$ \Msol \\
  \\
  \dxvii               & $-999\,\promille$  & $-998\,\promille$ & $-999\,\promille$  & $-997\,\promille$  & $-994\,\promille$ \\
  \dxviii              & $-1000\,\promille$ & $-996\,\promille$ & $-1000\,\promille$ & $-1000\,\promille$ & $-1000\,\promille$ \\
  $x$                  & 0.280            & 0.00241         & 0.514            & 0.0698           & 0.0587 \\
  \\
  Final \dxvii         & $-266\,\promille$ & $-62.3\,\promille$ & $-379\,\promille$ & $-121\,\promille$  & $-112\,\promille$ \\
  Final \dxviii        & $-266\,\promille$ & $-62.2\,\promille$ & $-379\,\promille$ & $-121\,\promille$  & $-112\,\promille$ \\
  Final $\Delta\oxvii$ & $-147\,\promille$ & $-30.6\,\promille$ & $-227\,\promille$ & $-61.3\,\promille$ & $-56.1\,\promille$ \\
  $\Delta$t 		& 1.51 Myr		& 0.77 Myr		& 1.03 Myr		& 1.34 Myr		& 1.75 Myr \\
\enddata
\label{tb:1Db}
\end{deluxetable}

  \begin{deluxetable}{lrrrrr}
 \tablewidth{0pt}
 \tablecaption{Oxygen isotopic shifts following injection from a 1D supernova}
 \tablehead{
 \colhead{$\,$}  & \colhead{16m-}    & \colhead{23m-}    & \colhead{40m-}    & \colhead{23e-}    & \colhead{23e-0.7-} \\
 \colhead{$\,$}  & \colhead{average} & \colhead{average} & \colhead{average} & \colhead{average} & \colhead{average} }
  \startdata
  \multicolumn{6}{c}{\underline{sub-explosive C burning }}\\
  \\
  \oxvi & $1.78\times10^{-1}$ \Msol     & $3.85\times10^{-1}$ \Msol     & $2.47$ \Msol  & $1.84$ \Msol  & $1.91$ \Msol \\
  \oxvii        & $1.81\times10^{-4}$ \Msol     & $1.03\times10^{-6}$ \Msol     & $1.88\times10^{-7}$ \Msol     & $4.18\times10^{-5}$ \Msol     & $4.86\times10^{-5}$ \Msol \\
  \oxviii       & $3.12\times10^{-4}$ \Msol     & $7.32\times10^{-6}$ \Msol     & $6.51\times10^{-8}$ \Msol     & $8.31\times10^{-5}$ \Msol     & $9.14\times10^{-5}$ \Msol \\
  \al   & $2.95\times10^{-6}$ \Msol     & $6.77\times10^{-5}$ \Msol     & $1.54\times10^{-5}$ \Msol     & $1.42\times10^{-5}$ \Msol     & $8.75\times10^{-6}$ \Msol \\
  \ca   & $1.65\times10^{-7}$ \Msol & $1.35\times10^{-6}$ \Msol & $8.42\times10^{-6}$ \Msol & $7.78\times10^{-6}$ \Msol & $8.44\times10^{-6}$ \Msol \\
  \\
  \dxvii               & $1503\,\promille$ & $-990\,\promille$ & $-1000\,\promille$ & $-944\,\promille$ & $-938\,\promille$ \\
  \dxviii              & $-221\,\promille$ & $-995\,\promille$ & $-1000\,\promille$ & $-980\,\promille$ & $-979\,\promille$ \\
  $x$                  & 0.0292          & 0.00276         & 0.0775           & 0.0628          & 0.106 \\
  \\
  Final \dxvii         & $-15.7\,\promille$ & $-62.6\,\promille$ & $-128\,\promille$  & $-112\,\promille$  & $-144\,\promille$ \\
  Final \dxviii        & $-64.6\,\promille$ & $-62.6\,\promille$ & $-128\,\promille$  & $-114\,\promille$  & $-148\,\promille$ \\
  Final $\Delta\oxvii$ & $+19.2\,\promille$  & $-30.7\,\promille$ & $-64.9\,\promille$ & $-55.3\,\promille$ & $-71.4\,\promille$ \\
 $\Delta$t 		& 0.90 Myr		& 0.72 Myr		& 1.29 Myr		& 1.29 Myr		& 1.39 Myr \\
\enddata
\label{tb:1Dc}
\end{deluxetable}
%==========================================================================================

GM07 likewise found, using the 1D models of \citet{Raus02}, that the ejecta are depleted 
in $\oxvii$, and in most cases $\oxviii$ as well. (Their $15 \, M_{\odot}$
case is enriched in $\oxviii$, in contrast to our $16 \, M_{\odot}$ case).
Similarly to our $23 \Msol$ models, both the $21 \Msol$ and the 
$25 \Msol$ model of \citet{Raus02} are very depleted in \oxvii~ and \oxviii, and the composition of 
those ejecta approach $-1000\, \promille$ for both \dxvii~ and \dxviii, although the \citet{Raus02}
models tend to be slightly richer in \oxvii~ and \oxviii~ than ours. The similarity between the \citet{Raus02} 
bulk abundances and those calculated for this work are unsurprising. Post-He burning stages rapidly 
destroy  $\oxvii$ and $\oxviii$, resulting in the oxygen in interior zones being nearly pure $\oxvi$. 
A 20-25 \Msol model has a very large oxygen mantle. The slightly higher fraction of heavy isotopes 
in the \citet{Raus02} arise in a somewhat larger He shell that results from a less accurate treatment 
of mixing in their earlier stellar models. 

These differences reflect the variability inherent in calculations of nucleosynthesis in 
massive stars, especially where small shifts in stable isotopes are concerned.
We also find, as did GM07, that the isotopic shifts associated with injection from ejecta from 
1D supernova explosions {\it tend} to be large (tens of permil), but not in all cases.
In the 23m runs, $\al$ is produced more abundantly, and the ${}^{26}{\rm Al} / {}^{16}{\rm O}$
ratios yield $x < 10^{-2}$ in these explosions.
In the 23m cases, the ejecta are particularly ${}^{16}{\rm O}$-rich, but relatively less oxygen
needs to be injected per $\al$ because more $\al$ is produced. (It should be remembered that 23m 
is a binary case, where a significant fraction of the $\oxvii$ and $\oxviii$ have been removed by 
mass loss from the He shell, and production of $\al$ has been enhanced by higher peak shock 
temperatures relative to the 23 $\Msol$ single star models.)
The isotopic shifts associated with injection from 23m 1D explosions are typically
$< 3\,\promille$.

For all the cases the shifts in both $\dxvii$ and $\dxviii$ are negative and similar in magnitude; that is, 
the injection of material from this supernova moves the composition down
the slope-1 line.
Later nebula evolution would produce materials that move back up this line. The most favorable case 
is the 23m supernova.
The magnitude and direction of the isotopic shifts associated with the 23m case
are such that current measurements of solar nebula materials do not rule out this possibility, even 
for bulk abundances. As we argue below, however, bulk abundances are not the best representation
 of the abundances of injected material.
%\begin{color}{red}
%Patrick, Carola, the referee really stuck on why this model is so different.  Can you explain
%why? 
%\end{color} 
%{\color{green}{
%the 23m models/cases have x $\sim$ 1e-3, 1-2 ord. of mag. smaller than the others, i.e. high Al compared to 
%oxygen. they also have high 16O compared to 17,18O. plus, it fits right with rauscher's 21 and 25 Msol 
%models, which have delta values going towards (-1000, -1000) also. so I still don't see what's so 
%"different" about it??
%}}
%\begin{color}{magenta} I think this is best dealt with in the response, reiterating to the referee that the 
%results are very similar.  The information is in the text. Three guesses on the referee's problem: Either 
%he is stuck on why the Al is different between the 23m and the 23e, he believes the model must be 
%different because the conclusions are different, which is really the result of the initial solar composition 
%vs. SMOW, or he's confusing the bulk results with the local results again. Are the larger shifts in other 
%models, say -100 - -200 per mil ruled out by the measurements of solar nebula materials? 
%\end{color}

Tables~\ref{tb:1Da}, \ref{tb:1Db}, and~\ref{tb:1Dc} and Figure \ref{fig:3isob} 
also show the isotopic shifts associated with 
injection from only $\al$-rich regions within the supernova.
Even when considering injection of $\al$-rich regions only, the conclusions are not much changed:
the isotopic shifts in oxygen generally are many tens of permil, and make the solar nebula more
${}^{16}{\rm O}$-rich. The sub-explosive C burning region of the 16m model is the only case that does 
not move the nebular composition along the slope 1 line. In general the sub-explosive C burning in the 
higher mass models provide the best results, as they produce the highest ratio of $\al$ to oxygen.

Tables~\ref{tb:1Da}, \ref{tb:1Db}, and~\ref{tb:1Dc} also give the yields of $\ca$ produced in
each of the 1D explosion scenarios.
The post-injection ${}^{41}{\rm Ca} / {}^{40}{\rm Ca}$ ratio is generally more than sufficient 
to match the meteoritic ratio, and a time delay is implied before isotopic closure, so that 
$\ca$ can decay. 
For the ejecta from the $23 \, M_{\odot}$ progenitors, the implied time delay (for $\ca$ to 
decay to a level ${}^{41}{\rm Ca} / {}^{40}{\rm Ca} = 1.4 \times 10^{-8}$) for the 23m cases is
$\sim$ 0.7 Myr for all three regions considered (i.e. bulk, explosive C/Ne burning, and sub-explosive 
C burning).
The implied time delay for injection from the 23e cases is $\sim$ 1.3 Myr for all three regions, 
and from the 23e-0.7 cases is 1.4 Myr (for the sub-explosive C burning region) -- 1.7 Myr (for the other 
two regions). The time delays for the 
other progenitor cases are all within those ranges. The range of these time delays are very similar to 
the range of 1.0 -- 1.8 Myr calculated by \citet{Goun07}.
The effect of this time delay is to cause $\al$ to decay, too, before isotopic closure, and 
to increase the isotopic shifts in oxygen.
The shifts are increased by factors of 2 (for the 23m) at the low end to 5.2 (for the 23e-0.7 cases) 
at the high end.
If isotopic closure is to be achieved in a few $\times (10^{5} - 10^6) \, {\rm yr}$ \citep{MacPD95, KitaH05},
then injection from the 23e and 23e-0.7 would seem to introduce too much 
$\ca$ to match constraints.
Injection from the more energetic 23m progenitor cases are consistent with a small shift in oxygen isotopes downward along the slope-1 line,
as well as the final ${}^{41}{\rm Ca} / {}^{40}{\rm Ca}$ ratio of the solar nebula.

\subsection{Asymmetric Supernova Explosions}

In Table~\ref{tb:3D} and Figure \ref{fig:3isob}, 
we present the yields of $\al$, $\ca$, and oxygen isotopes in various 
regions of the ejecta in our simulation of the 3D explosion.
We calculate the isotopic shifts if the injection uniformly samples all of the ejecta (bulk),
if it samples the Ring material, and if it samples the Bubble material.
By design, membership in the Bubble and Ring material is defined by high $\al$ content.
These $\al$-rich regions do not have well-defined edges, instead fading out monotonically
in $\al$-abundance as one moves out into the surrounding ejecta (see Figure \ref{fig:al}).
In order to not impose an arbitrary geometry on these regions we determined
membership by $\al$ amount per SNSPH particle. 
We used two different lower limits or thresholds for inclusion -- $1.5 \times 10^{-13} \, \Msol$ 
of $\al$ per particle for a maximum extent of the \al\, rich region, and 
$1.5 \times 10^{-11} \, \Msol$ per particle for a minimum extent of the $\al$-rich region 
(``high \al" case).
The most $\al$-rich SPH particles in the Ring and Bubble had $1.5 \times 10^{-10} \, \Msol$
and $4.8 \times 10^{-10} \, \Msol$ of $\al$, respectively. 
Each threshold picked out {\it all} SPH particles in the respective regions that it identified.

%==========================================================================================
% TABLE 3 - 3D
\begin{deluxetable}{lrrrrr} 
 \tablewidth{0pt}
 \tablecaption{\al\, and O in the 3D explosion \label{tb:3D}} 
 \tablehead{
 \colhead{$\,$}   & \colhead{Bulk}      & \colhead{Ring}       & \colhead{Ring}      & \colhead{Bubble}    & \colhead{Bubble} }
 \startdata                  
                  &                     &                      & high Al26           &                     &  
high Al26           \\
 \al\,(\Msol)      & $1.474\times10^{-6}$ & $3.510\times10^{-7}$  & $2.420\times10^{-7}$ & $9.996\times10^{-7}$ & $9.585\times10^{-7}$  \\
 \oxvi\,($\Msol$)  & 0.511               & $2.888\times10^{-2}$  & $5.949\times10^{-3}$ & $7.025\times10^{-3}$ & $2.550\times10^{-3}$  \\
 \oxvii\,($\Msol$) & $2.170\times10^{-4}$ &  $3.400\times10^{-5}$ & $2.241\times10^{-5}$ & $1.021\times10^{-5}$ & $1.738\times10^{-9}$  \\ 
 \oxviii\,($\Msol$) & $1.509\times10^{-1}$ &  $3.553\times10^{-3}$  & $1.018\times10^{-4}$ & $7.314\times10^{-3}$ & $5.635\times10^{-3}$  \\ 
 \ca\,(\Msol) & $2.132\times10^{-8}$ & $2.039\times10^{-8}$ & $4.312\times10^{-9}$ & $1.827\times10^{-11}$ & $1.239\times10^{-11}$ \\
                      \\
 $\delta\oxvii$   & $+43.9\,\promille$    &  $+1894\,\promille$    & $+8258\,\promille$    & $+2573\,\promille$    & $-998.3\,\promille$    \\
 $\delta\oxviii$  & $+129887\,\promille$  &  $+53546\,\promille$   & $+6582\,\promille$    & $+460545\,\promille$  & $+978646\,\promille$   \\ 
 $x$              & 0.16808             & 0.03987              & 0.01192             & 0.003406            & 0
.001289             \\
 \\
 Final $\delta\oxvii$ & $-45.1\,\promille$    &  $+14.8\,\promille$    & $+37.9\,\promille$    & $-51.1\,\promille$    & $-61.2\,\promille$     \\
 Final $\delta\oxviii$& $+18638\,\promille$   &  $+1995\,\promille$    & $+18.2\,\promille$    & $+1503\,\promille$    & $+1200\,\promille$     \\
 Final $\Delta\oxvii$ & $-1608\,\promille$    &  $-561\,\promille$     & $+27.8\,\promille$    & $-534\,\promille$     & $-477\,\promille$ \\
 $\Delta$t 		& 0.66 Myr		& 0.90 Myr		& 0.70 Myr		& --		& -- \\

\enddata
\end{deluxetable}
%==========================================================================================

Overall, the ejecta of the 3D simulation are much richer in \oxvii\, and \oxviii\, than the 1D
simulations, but also contain  two regions (the Ring and the Bubble) in which the $\al$
production is increased over the 1D calculations.
As we have previously discussed, the production of \oxviii\, is significantly altered
from the 1D results. 
The added yield from the decay of \fxviii\, to \oxviii\, makes the ejecta significantly richer 
in this isotope, and results in large (tens of permil) to very large
(hundreds of permil) and positive shifts in \dxviii\, for material from both the Ring and
Bubble, and the bulk. 
This is in stark contrast to the \oxviii\, poor ejecta produced in the 1D simulations, and emphasizes 
the sensitive dependence on the prevailing thermodynamic conditions of \oxviii\, production. 
The production of \oxvii\, is much less sensitive to the thermodynamic conditions.
In the 3D simulation we also see an increase over the 1D cases in the production of \oxvii\, in the 
\al\, rich regions and the bulk; however the change is not as drastic as in \oxviii. 
This again differs from the 1D calculations, and the more $\oxvii$-rich ejecta result in positive 
shifts in \dxvii, on the order of $-1 \, \rm{to }+15\,\promille$ for the Bubble and bulk,
and close to $+100\,\promille$ for the Ring.

Table~\ref{tb:3D} also shows the ${}^{41}{\rm Ca} / {}^{40}{\rm Ca}$ ratio following injection
of material from the 3D supernova into the solar nebula.
If injection comes from the Bubble region only, the amount of $\ca$ injected is too low to 
conform to meteoritic ratios, and injection from the Bubble can be ruled out on these grounds.
Injection of material from the Ring or bulk regions, in contrast, imply reasonable time delays 
$\approx 0.66 - 0.90 \, {\rm Myr}$. 
This implies an increase in oxygen isotopic shifts of $< 2.5$ over what is presented in Table~\ref{tb:3D}. 
These time delays are just below the ones \citet{Goun07} calculate, which again is explained by 
the faster density- drop off in the 3D calculation producing slightly less \ca~ than in 1D.

When an explosion samples a variety of thermodynamic trajectories through asymmetry, including those 
that result in freeze-out conditions due to rapid expansion, the overriding conclusion to be derived is that a very large range in 
oxygen isotopic shifts is allowed.  
It would seem extremely unlikely that conditions in an asymmetric explosion would conspire to 
yield a small isotopic shift consistent with the meteoritic constraints, though more``normal" trajectories that 
do not experience this freeze-out process are still candidate production sites, as wee see in 1D. 

%====================================================================%
% SECTION 5: DISCUSSION                                              %
%====================================================================%

\section{Discussion} 

As \citet{NichP99} strongly advocated, injection of supernova ejecta can produce measurable
``collateral damage" to stable isotope systems in protoplanetary disks.
GM07 in particular point to the role of oxygen isotopes in constraining this process.
The point of that paper was that the injection of $\al$ (and $\ca$) from a single nearby 
supernova necessarily would have brought in significant levels of oxygen isotopically distinct
from the pre-injection solar nebula.
The solar nebula after injection, they argued, would differ in its oxygen isotopes
by several tens of permil from the pre-injection values, which they robustly predicted
would be more ${}^{17}{\rm O}$-rich than the solar nebula.
They cited the {\it Genesis} measurements of solar wind oxygen as those most likely to
sample the pre-injection solar nebula.
Since preliminary results from {\it Genesis} \citep{Mcke09, }(McKeegan et al.\ 2009, 2010) 
are revealing the Sun to be ${}^{16}{\rm O}$-rich, GM07 would rule out injection
of $\al$ and $\ca$ from a single supernova. 

In this paper, we attempted to reproduce the calculations of GM07, to apply
their method of using oxygen isotopes to test the supernvoa injection hypothesis.
We made necessary corrections to their method, mostly in regard to the presumed
oxygen isotopic composition of the (post-injection) solar nebula.
GM07 assumed this was identical to SMOW, meaning the pre-injection 
solar nebula had to be more $\oxvii$-rich than almost any known inclusions.
We presented considerable evidence that the post-injection composition was in fact 
much more ${}^{16}{\rm O}$-rich than that, closer to $(-60\, \promille,-60\, \promille)$. 
We carried out stellar nucleosynthesis calculations, to calculate the isotopic yields
of $\al$, $\ca$ and oxygen isotopes in a variety of supernova explosion scenarios, 
including the 1D (spherically symmetric) cases 
as well as 3D (asymmetric) explosions. 
Because $\al$ and $\ca$ are observed to be correlated \citep{SahiG98}, 
we also simultaneously considered injection of $\ca$ into the solar nebula. 
We then computed the shifts in oxygen isotopes and the final ${}^{41}{\rm Ca} / {}^{40}{\rm Ca}$
ratio in the solar nebula following injection of sufficient supernova material to produce
the meteoritic ratio ${}^{26}{\rm Al} / {}^{27}{\rm Al} = 5 \times 10^{-5}$. 

Our 1D simulations largely confirm the results of GM07, that isotopic shifts are likely
to be tens of permil and to make the solar nebula more ${}^{16}{\rm O}$-rich than before
the injection.
We found that injection of material from either the bulk or the explosive C/Ne burning and sub-explosive C burning
regions of supernovae moved the composition of the solar nebula down the slope 1 line. Our 
$23 \Msol$ progenitors led to isotopic shifts in oxygen
 which moved the composition of the solar nebula down the slope-1 line, with the less energetic 
explosions producing larger shifts and time delays.
The 23m progenitors, which were the most energetic of the $23 \Msol$ cases and especially effective in
producing $\al$, generated shifts that amounted to only $< 6\, \promille$, including a time delay of 
0.7 Myr for $\ca$ to decay to its meteoritic value. 
This scenario, at least, is consistent with all the applied meteoritic constraints. If less than 100\% of the
 oxygen penetrated the solar nebula material due to, for example, dust condensation, all but one of the 
 1D cases are consistent with the evidence from the early solar system.

We note that this conclusion differs from what GM07 infer for injection of bulk ejecta from
21 and $25 \, M_{\odot}$ progenitors. 
GM07 likewise found isotopic shifts downward along the slope-1 line, but with a magnitude
of 40 to 50 permil.
It is worth noting that had GM07 assumed the same starting composition for the solar nebula,
$(-60\, \promille,-60\, \promille)$, 
that we do, 
then they would have found the solar nebula oxygen isotopic composition to be 
$(-82\, \promille,-82\, \promille)$ after injection of supernova material from an
$25 \, M_{\odot}$ progenitor, and $(-81\, \promille,-74\, \promille)$ after injection of supernova material from an
$21 \, M_{\odot}$ progenitor. Although these shifts are moderately large, they are {\it down} the slope-1 line.
As we established in \S \ref{magofshifts}, this would not have been {\it in}compatible with the meteoritic constraints, as some 
very ${}^{16}{\rm O}$-rich meteoritic samples in this range are known, including CAIs in 
Isheyevo, at $\approx (-68\,\promille,-66\,\promille)$ \citep{GounK09},
and a ferromagnesian cryptocrystalline chondrule in the CH chondrite Acfer 214,
at $\approx (-75\,\promille,-75\,\promille)$ \citep{KobaI03}.
Subsequently the mass-independent fractionation process would have shifted the nebula upward
along the slope-1 line, erasing this isotopic shift and eventually producing the composition
$(\dxvii,\dxviii) \approx (-40\,\promille,-40\,\promille)$ common to most CAIs
\citep[e.g.][]{ItohK04}.
We conclude that the supernova injection hypothesis cannot be rejected based
on 1D models. 

Our investigation of other parameters suggest that it is even more difficult to be conclusive
about supernova injection.
We have considered a small number of progenitor masses undergoing spherically symmetric collapse;
in a few cases we varied other parameters, such as varying the explosion energy, or allowing for loss of a
hydrogen envelope in a binary scenario, or allowing an asymmetrical explosion.
In most of these cases the isotopic shifts in oxygen were large.
Among the cases considered here, the final $\dxvii$ values in the solar nebula varied from
$-379\,\promille$ to $+15\,\promille$, and the final $\dxviii$ values varied from
$-379\,\promille$ to $+18000\,\promille$.
As GM07 found, most of the cases where meteoritic abundances of \al\, are injected lead to 
large ($>10\,\promille$) shifts in oxygen isotopes. 
We also considered the yields in a 3D anisotropic explosion of a $23 \, M_{\odot}$
progenitor, in the bulk ejecta and two $\al$-rich zones analogous to those in the 1D
explosions.
We find that a wide range of outcomes is possible, with oxygen isotopic shifts as large
as hundreds of permil, or as low as $< 3\,\promille$.
The fact that ${}^{18}{\rm F}$ can decay to ${}^{14}{\rm N}$ instead of ${}^{18}{\rm O}$
at high temperatures makes the yield of ${}^{18}{\rm O}$ especially sensitive to the
thermodynamic trajectory of the ejecta, which partially accounts for the spread in the
$\oxviii$ yields.
On the one hand, the wide range of possible outcomes makes it nearly impossible to state 
conclusively that all supernova injection scenarios can be ruled out.
On the other hand, the wide range of possible outcomes seems to imply a degree of fine
tuning so that the oxygen isotopic shifts in the solar nebula were not large, especially
for the 3D case.

We conclude that the hypothesis, that the $\al$ in the solar nebula was due to supernova 
material injected into the Sun's protoplanetary disk, can still be made compatible with
meteoritic constraints, under two scenarios.
The first is that the injected supernova material came from either the bulk ejecta, or
from 
a region in a supernova that experienced thermodynamic conditions like the 
subexplosive C burning zone. The latter is more physically likely.
With a $0.7$ Myr time delay, the injection would have moved the solar nebula oxygen
isotopic composition from 
$(\dxvii,\dxviii) \approx (-60\, \promille,-60\, \promille)$ to a more $\oxvi$-rich value along the slope one 
line. All but one of our explosions produce movement along the slope 1 line. We produce shifts as small as 
$\approx (-63\, \promille,-63\, \promille)$, which would have produced an accompanying meteoritic ratio  
${}^{41}{\rm Ca} / {}^{40}{\rm Ca} = 1.4 \times 10^{-8}$.
Subsequent mixing of rocky material with a ${}^{16}{\rm O}$-depleted reservoir would have
then moved the composition of meteoritic inclusions up the slope-1 line to values 
$(\dxvii,\dxviii) \approx (-50\, \promille,-50\, \promille)$, consistent with primitive
CAIs, and further up the slope-1 line with time.

The second scenario is one in which only dust grains are injected in to the protoplanetary
disk, and very little of the supernova oxygen condenses into dust grains. 
If only the most refractory grains such as corundum were injected, then potentially
$x \ll 10^{-5}$, and the isotopic shifts would be negligible ($\ll 1\, \promille$),
for nearly all the cases considered here. 
It is worth noting that Ca is equally refractory to Al and is likely to condense from
supernova ejecta under the same conditions that Al condenses, so the meteoritic abundance
of $\ca$ could still be matched following injection of $\al$.
\citet{Ouel07} have calculated that only $1\%$ of gas-phase ejecta are injected 
into a disk.
If almost all of the Ca and Al in the ejecta are locked up in large grains 
(radii $> 0.1 \, \mu{\rm m}$) that are efficiently injected \citep{OuelDH10}, but
less than a few percent of the oxygen is, then 
potentially all of the isotopic shifts in oxygen calculated here should be reduced by a 
factor of about 100. 
Essentially all of the 1D cases considered here would then conform with the meteoritic 
constraints, and even some of the 3D cases as well.

To summarize, we agree with GM07 that oxygen isotopes can be a powerful constraint on 
supernova injection models.
Our calculations of oxygen isotopic shifts following injection from the bulk ejecta
of 1D supernovae broadly match the results of GM07.
Had GM07 assumed the same starting composition of the solar nebula that we did, and
considered a smaller time delay between injection and isotopic closure, they would
have found isotopic shifts for $20 - 25 \, M_{\odot}$ progenitors that would not
be inconsistent with meteoritic constraints.
Our own calculations of the same case predict shifts that are similar, although smaller
in magnitude, and which are also consistent with meteoritic constraints. 
The existence of an example that is consistent with the oxygen isotopic composition and
the ${}^{41}{\rm Ca} / {}^{40}{\rm Ca}$ ratio of the solar nebula means that the 
supernova injection hypothesis cannot be ruled out. 
Because the nucleosynthesis of oxygen differs in asymmetric explosions, a much wider
range of oxygen isotopes is possible in 3D explosions.  
Because of the contingent nature of the injection it becomes difficult to make any statement 
about the possibility that the solar nebula acquired
$\al$ from such an asymmetric explosion.
Finally, all oxygen isotopic shifts are reduced if only large grains are injected
into the protoplanetary disk, and only a small fraction of oxygen condenses into
large grains.
Quantifying the fractionation of Al and O during injection into a protoplanetary disk
is the focus of ongoing work by this research group.
If only a few percent of the total oxygen is injected, then nearly all the 1D explosions 
considered here could be consistent with the meteoritic constraints on oxygen isotopes
and $\ca$ abundances. 
We therefore conclude it is premature to rule out the supernova injection hypothesis
based on oxygen isotopes.

{}

\end{document}